\documentclass[aip,cha,graphicx]{revtex4-1}
\usepackage{epsfig}
\draft 

\begin{document}


\title{Peculiarities of wave fields in nonlocal media} 



\author{V.A.Danylenko}
\author{S.I.Skurativskyi}
\email[]{skurserg@gmail.com}
\affiliation{Division of geodynamics of explosion, Subbotin institute of Geophysics, NAS of Ukraine}


\date{\today}

\begin{abstract}
The article summarizes the  studies of wave fields in structured
non-equilibrium media describing by means of nonlocal hydrodynamic
models. Due to the symmetry properties of models, we derived the
invariant wave solutions satisfying  autonomous dynamical systems.
Using the methods of numerical and qualitative analysis, we have
shown that these systems possess  periodic, multiperiodic,
quasiperiodic, chaotic, and soliton-like solutions. Bifurcation
phenomena caused by the varying of nonlinearity and nonlocality
degree are investigated as well.
\end{abstract}



\maketitle 

\begin{quotation}
In order to describe non-equilibrium media when the manifestations
of intrinsic structure can not be ignored, we use hydrodynamic
mathematical models. Information about relaxing processes and
interactions between structural elements is incorporated in  the
dynamical equations of state (DES) which,  unlike the local
classic relations, now become nonlocal one in time and space.
Using the symmetry reduction scheme, we obtain the autonomous
dynamical systems describing the invariant wave solutions. By
qualitative analysis methods,  we  show that  the dynamical
systems possess periodic, multiperiodic, and chaotic solutions
obeying the Feigenbaum scenario. We study  the quasiperiodic
regimes and their bifurcations. We also reveal the existence of
homoclinic trajectories of Shilnikov type and investigate the
changes of  homoclinic structures when the bifurcation parameters
vary. Hidden attractors, hysteretic phenomena are discovered as
well. As a result, depending on the chosen model for media, we
classify the wave solutions and their bifurcations and show that
spatio-temporal nonlocal models are promising  for the describing
of complicated wave regimes in structured media.
\end{quotation}

\section{Introduction}

Open thermodynamic systems attract attention of scientists by
their synergetic properties, their ability to produce localized
nontrivial structures and order. Description of such phenomena
demands the  creation of new and the refinement of already known
mathematical models.

 According to Refs.~\onlinecite{VDK,DanDan,DanDanSkur}, using the methods of
non-equilibrium thermodynamics and the internal variables concept
\cite{DSV93}, the nonlinear temporally and spatially nonlocal
mathematical models for non-equilibrium processes in media with
structure have been constructed. In this report we present the
results of investigations of wave processes in such media. To do
this, we use the following hydrodynamic type system
\begin{eqnarray}
\displaystyle  \dot \rho  + \rho u_x =0, \quad
 \displaystyle \rho \dot u + p_x
 = \gamma \rho^m,\nonumber\\
 \frac{1}{{\rho ^2 }}\frac{{\Gamma \varepsilon _r }}{{\tau _{{\rm T}{\rm P}} }}
 \left\{ {\left[ { - \rho _{xx} \left( {1 + {\bf a}} \right) + \frac{1}{\rho }
 \left( {\rho _x } \right)^2 \left( {1 - {\bf a}\Gamma _{{\rm V}0} } \right)} \right]}
 \right. + \left[  - \ddot \rho  \right.(1 + \bf a) + \nonumber \\
 + \frac{2}{\rho } \dot \rho  ^2
 \left( {1 - \frac{{{\bf a}\left( {\Gamma _{{\rm V}0}  - 1}
 \right)}}{2}} \right) + \left. {\left. {\frac{1}{{\tau _{{\rm T}{\rm P}} }}
 \dot\rho \left( {1 + {\bf a}} \right)} \right]} \right\} + \omega _0^2
 \rho _0^{1 - \Gamma _{{\rm V}0} } \rho _{}^{\Gamma _{{\rm V}0} }  -  \label{ds:bal_eq}\\
  - \omega _0^2 \rho _0  = b\left( {p - p_0 } \right) +
  b\tau _{{\rm T}{\rm V}} \dot p - \frac{{\chi _{{\rm T}0} }}{{\chi _{{\rm T}\infty } }}
  b\tau _{{\rm T}{\rm V}}^2 \ddot p  - b\Gamma \varepsilon _r \tau _{{\rm T}{\rm V}} \left( {p_{xx}  + \frac{{\rho _x }}{\rho }p_x } \right),\nonumber
\end{eqnarray}
where
\begin{eqnarray*}
{\bf a} = T_0 \alpha _\infty  \Gamma _{{\rm V}0} \left(
{\frac{\rho }{{\rho _0 }}} \right)^{\Gamma _{{\rm V}0}  + 1}
,\quad \omega _0^2 = \frac{{bc_{S0}^2 \alpha _0 T_0 }}{{\gamma _0
}},\\
 b=\frac{\chi_{T0}}{\rho_0\tau_{TP}^2},\quad \chi _{{\rm T}0 } =
\rho_0^{-1} c_{{\rm T}0}^{-2} = \gamma _\infty  \rho_0^{-1} c_{S0
}^{-2};
\end{eqnarray*}
$ c_{T0}$, $c_{S0} $ are the isothermal and adiabatic frozen
velocities of sound; $ \gamma _\infty $ is the frozen polytropic
index.

Using the characteristic quantities $t_0,\,u_0,\,\rho_0$, let us
construct the scale transformation
\begin{eqnarray}
t=\bar t t_0 ,\quad x=\bar x t_0 u_0 ,\quad p=\bar p \rho_0 u_0^2,
\quad
\rho=\bar \rho \rho_0,\quad u=\bar u u_0  \nonumber\\
\displaystyle \sigma=\frac{\Gamma \varepsilon_r \tau_{TV}}{(t_0
u_0)^2},\quad  \tau_{pT}=\tau_{TV}\frac{\chi_{T0}}{\chi_{T\infty}}
,\quad \tau=\frac{\tau_{TV}}{t_0},\quad
h=\frac{\chi_{T0}}{\chi_{T\infty}}\tau^2,  \label{obez}
\\
\kappa=\frac{\omega_0^2}{b u_0^2} = \alpha_0 T_0 \gamma_0
\left(\frac{C_{T0}}{u_0}\right)^2 ,\quad \chi=\frac{1}{\rho_0
u_0^2\chi_{T\infty}} \equiv
\left(\frac{C_{T \infty}}{u_0}\right)^2,  \nonumber\\
  a=\delta n \rho^{n+1},\quad \delta
= T_0 \alpha_{\infty} , \quad \Gamma_{V0}=n,\nonumber
\end{eqnarray}
which leads  system (\ref{ds:bal_eq})  to the dimensionless form
\begin{equation}\label{state}
\begin{array}{c}
\displaystyle  \dot \rho  + \rho u_x =0, \quad
 \displaystyle \rho \dot u + p_x
 = \gamma \rho^m, \\
\displaystyle   \sigma \chi \rho^{-2}
\left[-\rho_{xx}(1+a)+\rho_x^2 \rho^{-1}(1-a n)\right]+
\\
\displaystyle +h\, \chi\,\rho^{-2}\left[-\ddot \rho(1+a)+2\dot
\rho^2 \rho^{-1}(1-0.5a(n-1))+ \tau h^{-1}\dot\rho(1+a)\right] +
\\  \displaystyle
+\kappa \rho^n= p+\tau \dot p - h \ddot p
 \displaystyle -\sigma \left(p_{xx}+\rho_x p_x \rho^{-1}\right).
\end{array}
\end{equation}
We would like to emphasize that system (\ref{state}) can be
regarded as an hierarchical set of submodels which are complicated
by means of taking new effects into account. We thus  are going to
study the chain of nested models  and classify their wave
solutions using the methods of qualitative and numerical analysis.

The remainder of the  report is organized as follows. In
Sec.\ref{sec2} we begin our studies from the simplified version of
system (\ref{state}) keeping the terms with the first temporal
derivatives, then attaching the terms with the second  temporal or
spatial derivatives. The form of wave solutions and the
description of techniques for their exploration are presented in
detail. Sec.\ref{sec3}  is devoted to the spatially nonlocal model
 which is used for investigating of the
Shilnikov homoclinic structures whose existence and bifurcations
are extremely important during chaotic regimes formation. The
model incorporated both temporal and spatial nonlocalities are
presented in Sec.\ref{sec4}. Generalizations of the previous
models by means of introducing  the third temporal derivatives and
incorporating of physical nonlinearity are given in Sec.\ref{sec5}
and Sec.\ref{sec6}, respectively. For all models we derive
invariant wave solutions and carry out the qualitative analysis of
 corresponding factor-systems.

\section{Wave solutions of the models with DES incorporating the
second temporal or spatial derivatives}\label{sec2}

To begin with, let us consider the simplest model with relaxation
derived  from (\ref{state}) at $\delta=h=\sigma=0$, $n=1$. As has
been shown in Refs.~\onlinecite{DSV93,vsan_op}, the system
\begin{equation}\label{opusc}
  \dot \rho  + \rho u_x =0, \quad
\rho \dot u + p_x
 = \gamma \rho, \quad
\tau(\dot p-\chi \dot\rho)=\kappa\rho-p,
\end{equation}
due to its symmetry properties \cite{symm}, admits the ansatz
\begin{equation}\label{wave_sol}
u=U(\omega)+D,\,\rho=\rho_0\exp(\xi t+S(\omega)),\, p=\rho
Z(\omega),\, \omega=x-Dt,
\end{equation}
where $D$ is the constant velocity of wave front, $\xi$ determines
a slope of the inhomogeneity of the steady solution
(\ref{wave_sol}). According to Ref.~\onlinecite{vsan_op},
solutions (\ref{wave_sol}) are described by the plane system of
ODE which possesses  limit cycles  and homoclinic trajectories.

If    we incorporate the second temporal derivatives in the last
equation of system (\ref{state}), then the previous DES is
generalized to the following one:
\begin{equation}\label{temporal2}
\tau \left( {\frac{dp}{dt} - \chi \frac{{d\rho }}{dt}} \right) =
\kappa \rho  - p  - h\left\{ {\frac{{d^2 p}}{{dt^2 }} + \chi
\left( {\frac{2}{\rho }
  \left( {\frac{{d\rho }}{{dt}}} \right)^2  - \frac{{d^2 \rho }}{{dt^2 }}} \right)} \right\}.
\end{equation}
This model takes into account  the dynamics of internal relaxation
processes in more detail. As has been shown in
Ref.~\onlinecite{SidVlad97},  wave solutions (\ref{wave_sol}) are
described by the system of ODE with three dimensional phase space.
This system possesses the limit cycles  undergoing  the period
doubling cascade, and chaotic attractors.

Consider now the model with relaxation and spatial nonlocality
\begin{equation}\label{spatial2}
\tau\left(\dot p - \chi \dot \rho\right)=\kappa \rho -p +\sigma
\biggl\{  p_{xx}+\frac{1}{\rho}  p_{x}\rho_{x}-\chi\left(
\rho_{xx}- \frac{1}{\rho}\biggl( \rho_{x}\biggr)^2\right)\biggr\}.
\end{equation}
Solutions (\ref{wave_sol}) satisfy the following dynamical system
\begin{equation} \label{skur4}
\begin{array}{l}
\vspace{3mm} \displaystyle U\frac{dU}{d\omega}=UW,\qquad
 U\frac{dZ}{d\omega}=\gamma U+\xi Z+W(Z-U^2) , \\
\vspace{3mm} \displaystyle
 U\frac{dW}{d\omega} = \{ U^2 [\tau (\gamma U + \xi Z - WU^2 ) + \chi \tau W + Z - \kappa ] +  \\
 + \sigma [(\xi  + W)(2U(\gamma  - UW) + \chi W) + (UW)^2 ]\} \left[ {\sigma (\chi  - U^2 )} \right]^{ -
 1}.
 \end{array}
\end{equation}
This system has the fixed point
\begin{equation}\label{skur5}
U_{0}=-D,\quad  Z_{0}=\frac{\kappa}{1-2\sigma (\xi /D)^2},\quad
W_{0}=0, \quad \gamma=\xi Z_0/D
\end{equation}
which is the only one lying in the physical parameter range.

\begin{figure}[th]
\includegraphics[width=7.5 cm, height=6 cm ]{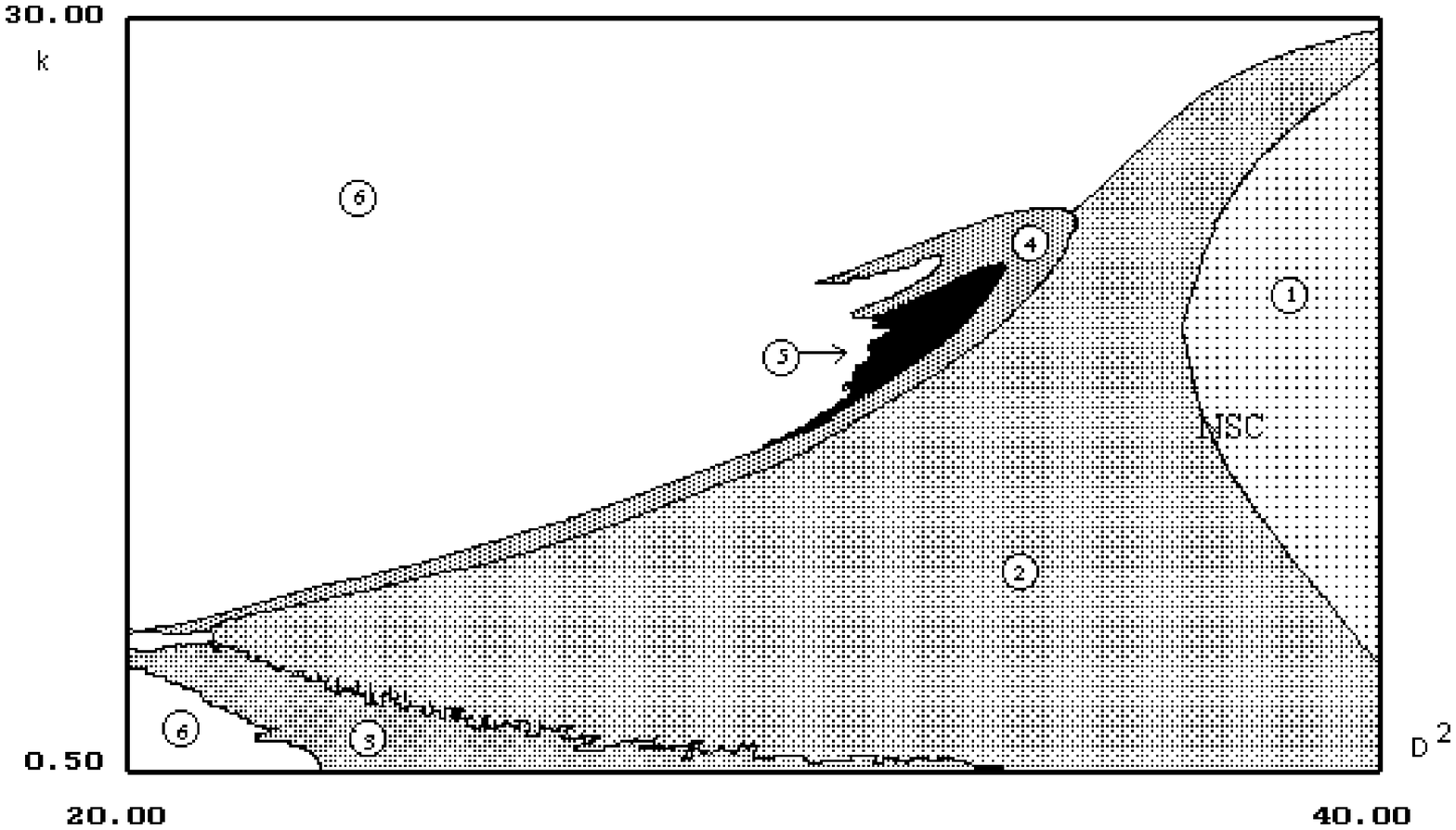}
\hfill
\includegraphics[width=7.5 cm, height=6 cm ]{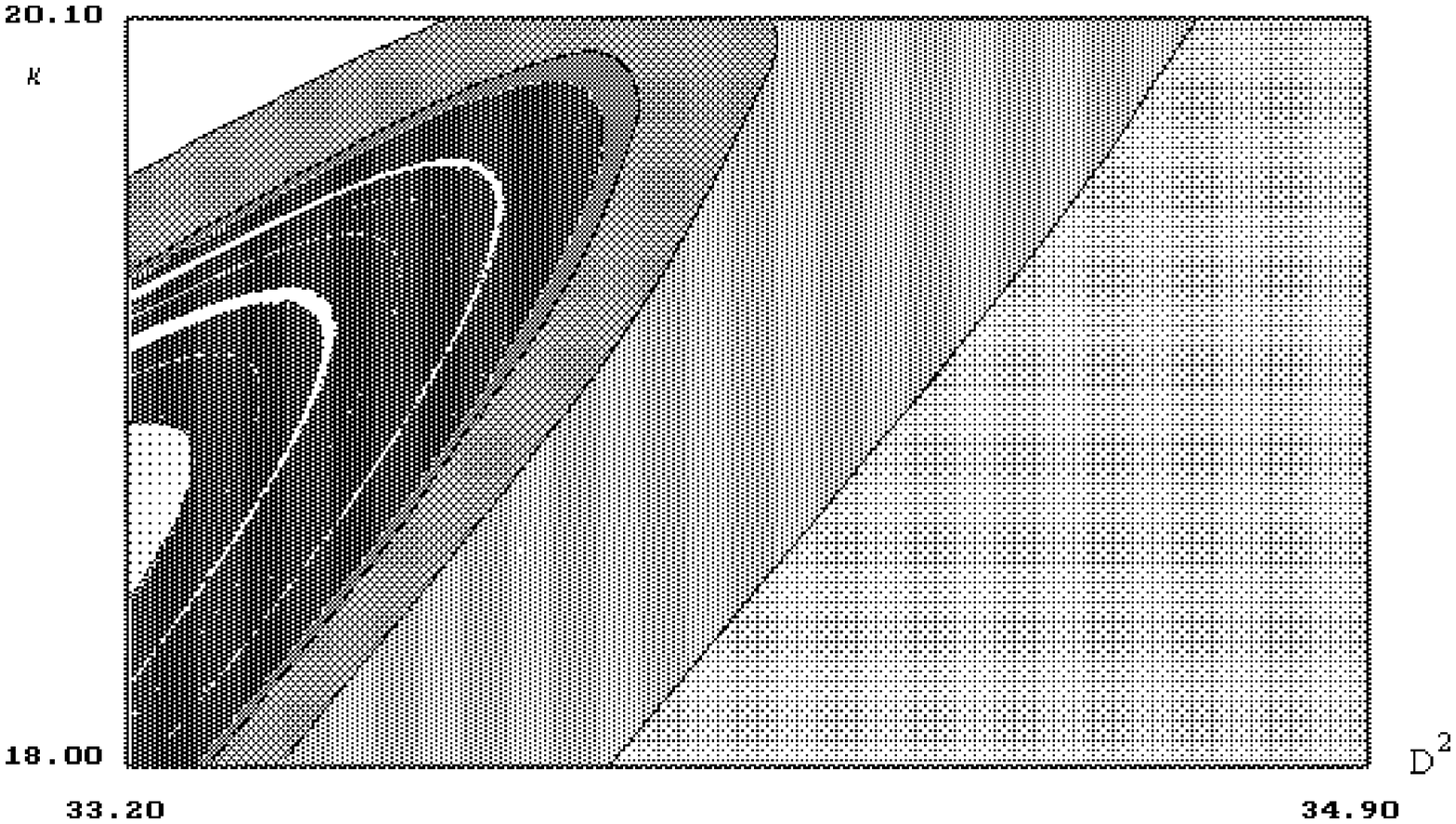}
 \centerline{a \hspace{8cm} b}
\caption{Left: bifurcation diagram of system (\ref{skur4}) in
parametric space $(D^2,\kappa)$: 1 -- stable focus; 2 --
$1T$-cycle; 3 -- torus; 4 -- multiperiodic attractor; 5 -- chaotic
attractor; 6 -- loss of stability. Right: enlargement of  part of
the left figure: 6 -- $3T$-cycle. }\label{spat_zag}
\end{figure}

We start from analyzing the linearized in the fixed point
(\ref{skur5}) system (\ref{skur4}) with the  matrix
 $\hat M$
$$\hat{M} = \left(\begin{array}{ccc} 0& 0 & -D\\ \gamma& \xi&
Z_0-D^2\\A& B& C \end{array} \right),$$ where
$$
A=\frac{D\kappa \xi ( 2\xi \sigma - D^{2}\tau)}  {Q\sigma (2\xi^2
\sigma-D^2 )}, \quad B=\frac{D^{2}(1 + \xi \tau )}{Q},\quad
Q=\sigma (\chi-D^2),
$$
$$
C=Q^{-1}\left\{ \xi \sigma \left(\chi-D^2\right)-
\frac{2D^{2}\kappa\xi\sigma}{D^2 - 2\xi^{2} \sigma} +
D^{2}\tau\left(\chi-D^2\right)\right\}.
$$
The well-known Andronov-Hopf bifurcation theorem \cite{GH} tells
us that periodic solution creation can take place if the spectrum
of matrix $\hat M$ looks as $(-\alpha;\pm \Omega i)$. This is so
if the following relations hold:
\begin{equation}\label{skur6}
    \alpha = \xi+C>0,
\end{equation}
\begin{equation}\label{skur7}
    \Omega^2 =AD-B\left(Z_0-D^2\right) +\xi C>0,
\end{equation}
\begin{equation}\label{skur8}
    \alpha \Omega^2 = \xi \left(AD-Z_0B\right)>0.
\end{equation}
The first two take on the form of inequalities imposing some
restrictions on the parameters. The third one determines the
neutral stability curve (NSC) in space $\left(D^2;\kappa\right)$
provided that the remaining parameters are fixed. For
$\sigma=0.76$, $\xi=1.8$, $\tau=0.1$, $\chi=50$, it looks like a
parabola with branches directed from left to right, see
Fig.\ref{spat_zag}a. Crossing the NSC from right to left, we
observe the limit cycle appearance. Development of limit cycle at
decreasing $D^2$ it is convenient to study by means of the
Poincar\'{e} section technique \cite{holodn,ds_rep_07}.

\begin{figure}[th]
\includegraphics[width=7.5 cm, height=6 cm ]{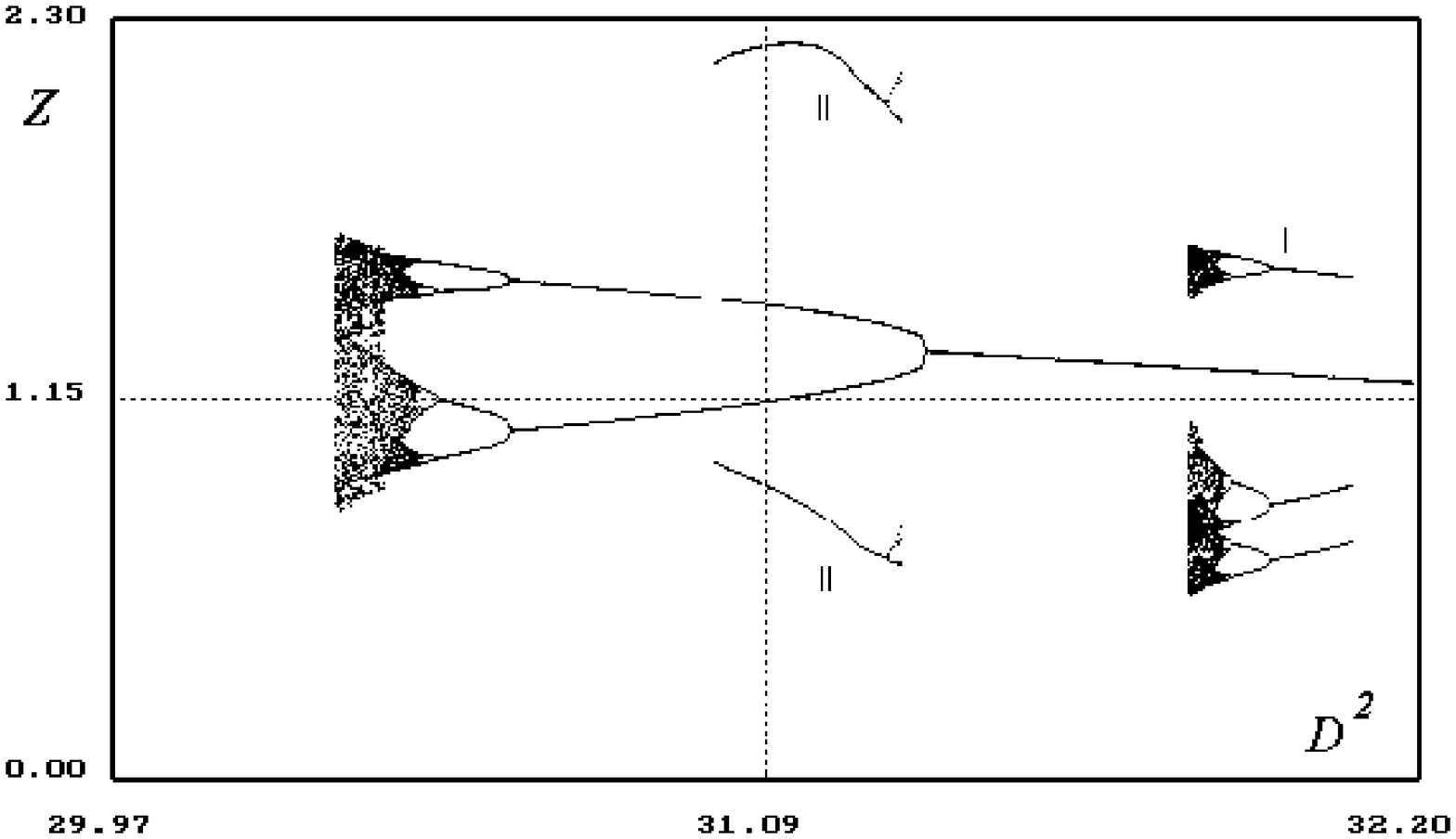}
\hfill
\includegraphics[width=7.5 cm, height=6 cm ]{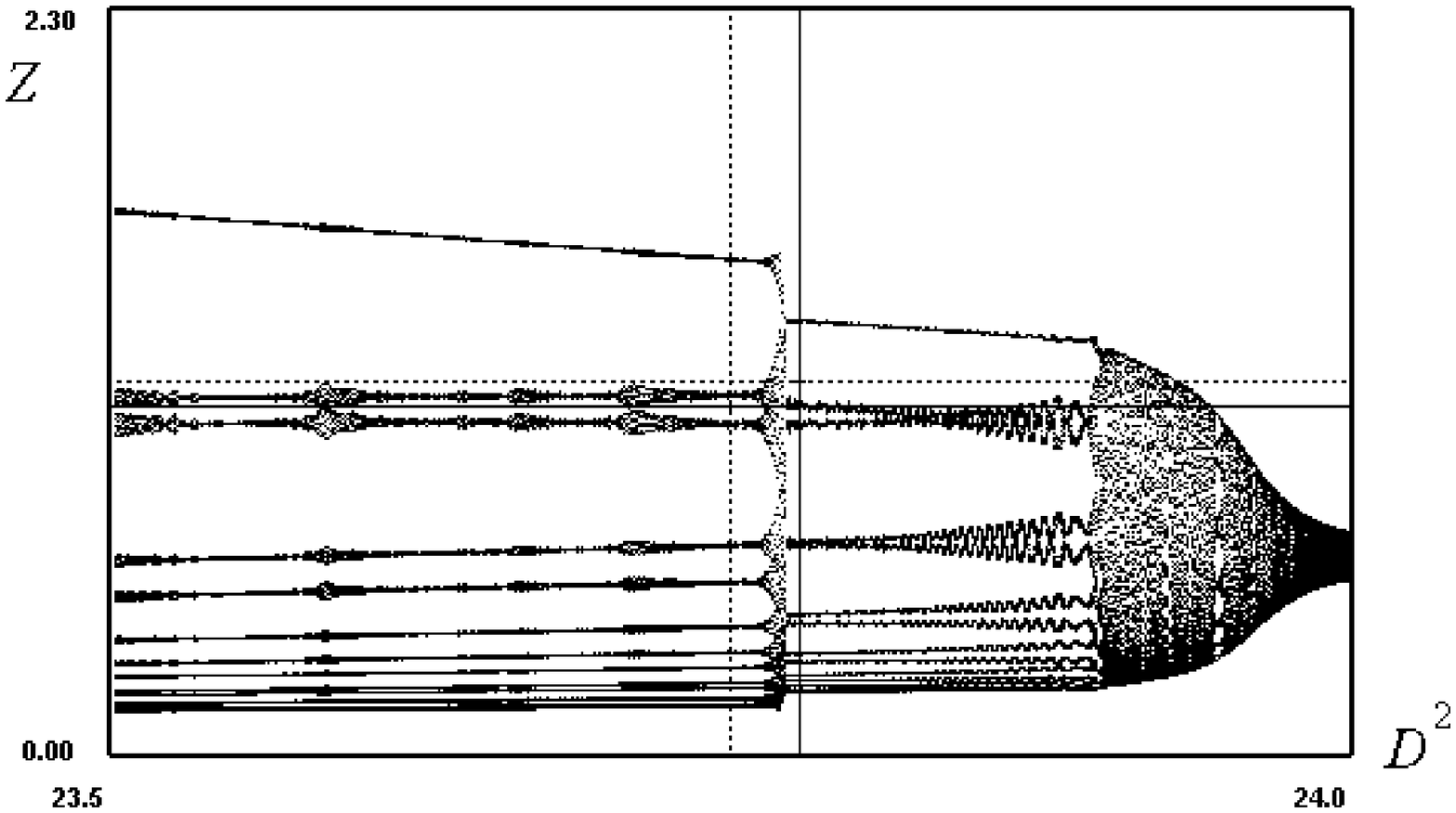}
 \centerline{a) \hspace{8cm} b)}
\caption{Bifurcation diagrams of system (\ref{skur4}) in plane
$(D^2,Z)$, obtained for  $\chi=\eta=50,\, \xi=1.8,\,\tau=0.1,\,
\sigma=0.76$ and (a) $\kappa=14$, (b) $\kappa=1$.
}\label{chaos_tor}
\end{figure}

Let us choose the plane $W=0$ as an intersecting one and find
coordinates of intersection points of phase curves which
cross-sect the intersecting plane only in one direction. Plotting
coordinate $Z$ of the cross-section point along the vertical axis,
and the value of the  bifurcation parameter $D^2$ along the
horizontal one, we will obtain the typical  bifurcation diagrams
(Fig.\ref{chaos_tor}). From the analysis of  diagram
Fig.\ref{chaos_tor}a we can see that while parameter $D^2$
decreases the development of the limit cycle coincides with the
Freihenbaum scenario, followed by the creation of a chaotic
attractor. Moreover, in the vicinity of the main limit cycle there
are the hidden attractors (depicted in Fig.\ref{chaos_tor}a by the
symbols I and II). These attractors can be visualized by the
integrating of  system (\ref{skur4}) with special initial data
only.

In Fig.\ref{chaos_tor}b we see the torus development at decreasing
$D^2$. According to the diagram, we can distinguish  tori with
densely wound trajectories and striped tori.

Doing in the same way, we get the two-parameter bifurcation
diagram (Fig.\ref{spat_zag}) which tells us that system
(\ref{skur4}) possesses the periodic, multiperiodic,
quasiperiodic, and chaotic trajectories.

Such a complicated structure of the phase space of the system can
be coursed by  homoclinic trajectory existence.

\begin{figure}[t]
\includegraphics[width=7.5 cm, height=6 cm ]{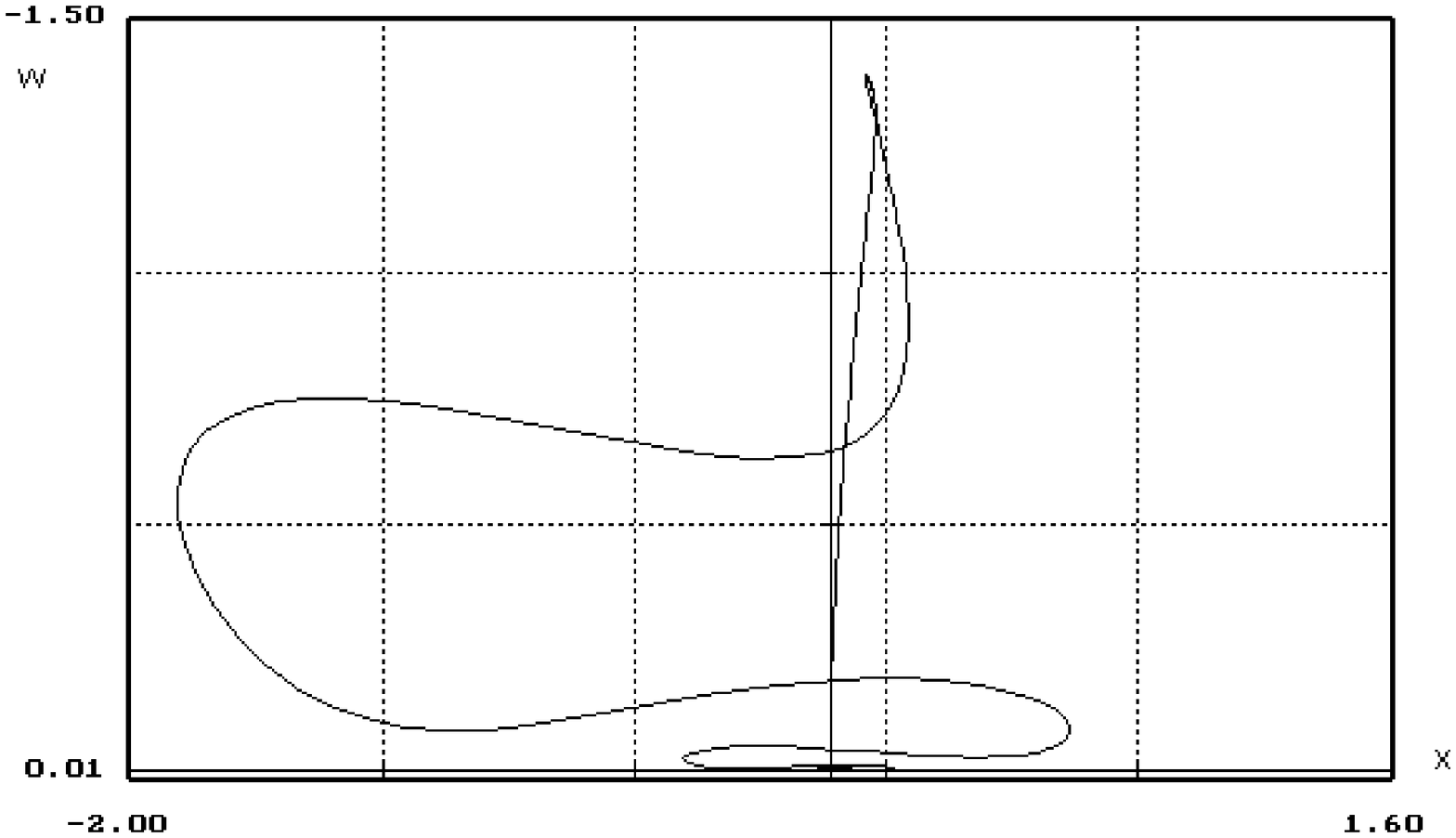}
\hfill
\includegraphics[width=7.5 cm, height=6 cm ]{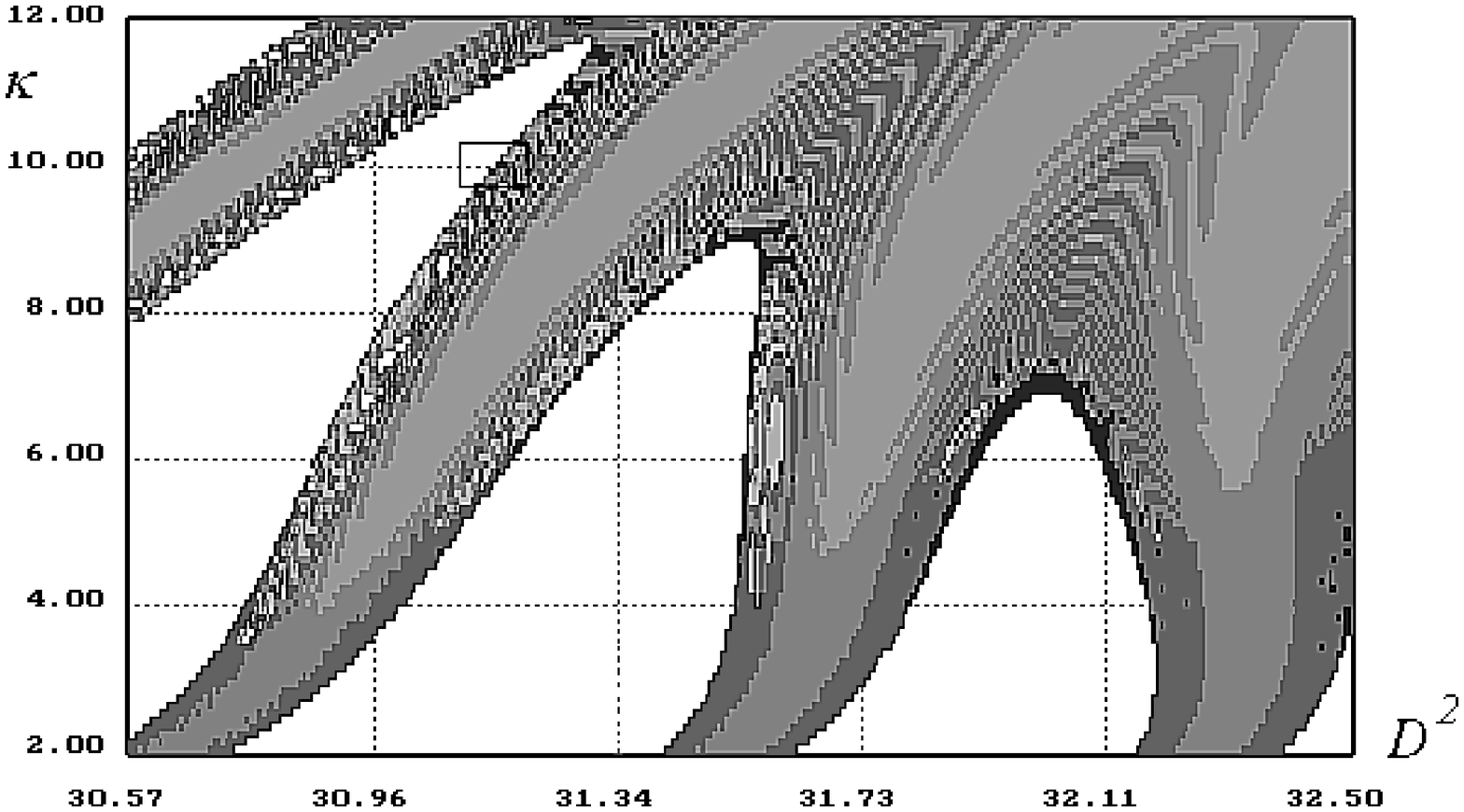}
 \centerline{a) \hspace{8cm} b)}
\caption{ a) Projection of the homoclinic solution of system
(\ref{skur4}) onto the $(X,W)$ plane. b) A portrait of subset of
parameter space
  $(D^2,\,\kappa)$, corresponding to different intervals of
  function $f^{\Gamma}_{\min}(D^2,\,\kappa)$ values and following
  Cauchy data: $X(0)=Y(0)=0,\,W(0)=0.001$: $f^{\Gamma}_{\min}>1.2$
  for white colour; $0.6<f^{\Gamma}_{\min}\leq 1.2$ for light grey;
  $0.3<f^{\Gamma}_{\min}\leq 0.6$ for grey;
  $0.01<f^{\Gamma}_{\min}\leq 0.3$ for deep grey;
  $f^{\Gamma}_{\min}\leq 0.01$ for black. }\label{phas_soliton}
\end{figure}



\section{Homoclinic loops of Shilnikov type and their
bifurcations}\label{sec3}

First it worth noting that existence of homoclinic trajectories,
i.e.  loops consisting of the separatrix orbits of hyperbolic
fixed point, plays a crucial role \cite{Butenin,homocl_wiggins} in
the formation of localized regimes in the phase space of dynamical
system.

For the present, the question on the existence of homoclinic
trajectory of Shilnikov type\cite{GH,kuznetsov}  in system
(\ref{skur4}) has been treated numerically.

We investigate a set of points of parameter space $(D^2,\,
\kappa)$ for which the trajectories moving out of the origin along
the one-dimensional unstable invariant manifold $W^u$ return to
the origin along the two-dimensional stable invariant manifold
$W^s$. In practice, for the given values of parameters $\kappa$,
$D^2$, we
 numerically define a distance (the counterpart of split function in Ref.~\onlinecite{kuznetsov}, p.198) between the origin and point
$(X^\Gamma (\omega),Y^\Gamma (\omega),W^\Gamma (\omega))$ of the
phase trajectory  $\Gamma\left(\cdot;\,\,\kappa,\,\,D^2\right)$:
$$
f^{\Gamma}\left(
\kappa,\,\,D^2;\,\,\omega\right)=\sqrt{\left[X^{\Gamma}(\omega)\right]^2+\left[Y^{\Gamma}(\omega)\right]^2+\left[W^{\Gamma}(\omega)\right]^2},
$$
starting from the fixed Cauchy data $(0,0,0.001)$. Next we
determine
\begin{equation}
\Phi(\kappa,\,\,D^2)=\min _ {\omega} \{
f^{\Gamma}\}\end{equation} for the part of the trajectory which
lies beyond the point at which the distance gains its first local
maximum, providing that it still lies inside the ball centered at
the origin and having a fixed (sufficiently large) radius (for
this case $f^\Gamma(\omega)\leq 5$). The results are presented in
Figs.\ref{phas_soliton}b. The first is of the most rough scale
among this series. Here, white color marks the values of
parameters $\kappa$, $D^2$ for which $\Phi>1.2$, light grey
corresponds to the cases when $0.9 <\Phi < 1.2$ and so on (further
explanations are given in the subsequent captions). The black
coloured patches correspond to the case when $\Phi < 0.01$. In
Ref.~\onlinecite{Vlad_Skur00} the structure of the set of points
from Fig.~\ref{phas_soliton}b  has been studied in more detail.


\section{Models with DES taking  spatial and temporal nonlocalities into account}\label{sec4}

Combining the model (\ref{temporal2}) and (\ref{spatial2}), we
obtain the following spatio-temporal nonlocal model
\begin{equation}\label{SPTEM2}
\tau\left(\dot p - \chi \dot \rho\right)=\kappa \rho -p +\sigma
\biggl\{  p_{xx}+\frac{1}{\rho} p_{x}\rho_{x}-\eta\left(
\rho_{xx}- \frac{1}{\rho}\biggl( \rho_{x}\biggr)^2\right)\biggr\}-
\\ h \biggl\{\ddot p+\eta \left( \frac{2}{\rho}\displaystyle (\dot
\rho)^2-\ddot \rho\right)\biggr\}.
\end{equation}
This model  has been studied in
Refs.~\onlinecite{VDS2004,skur_NPCS2001}, when the parameters $h$
and $\sigma$ are regarded as a small one, i.e., Eqs.~
(\ref{temporal2}) and (\ref{spatial2}) are perturbed by the terms
with high derivatives. It turned out that the wave localized
regimes are saved under perturbations and undergo some smooth
changes.

\section{Models involving  DES with the third temporal  derivatives}\label{sec5}

If we need to describe the relaxing processes in more detail, then
we can incorporate the terms with the third temporal derivatives
in DES (\ref{SPTEM2}). In this case DES has the form
\cite{DanDanSkur}
\begin{equation}\label{vizn1}
\begin{array}{l}
\displaystyle
  \tau \left( {\frac{{dp}}{{dt}} - \chi \frac{{d\rho }}{{dt}}} \right) = \kappa \rho  - p + \sigma
  \left\{ {\frac{{\partial ^2 p}}{{\partial x^2 }} + \frac{1}{\rho }\frac{{\partial p\partial \rho }}
  {{\partial x\partial x}} - \chi \left( {\frac{{\partial ^2 \rho }}{{\partial x^2 }} - \frac{1}{\rho }
  \left( {\frac{{\partial \rho }}{{\partial x}}} \right)^2 } \right)} \right\} -\\
 \displaystyle  {\rm{           }} - {\rm{ }}h\left\{
{\frac{{d^2 p}}{{dt^2 }} + \chi \left( {\frac{2}{\rho }
  \left( {\frac{{d\rho }}{{dt}}} \right)^2  - \frac{{d^2 \rho }}{{dt^2 }}} \right)} \right\} +
  \frac{{h^2 }}{\tau }\frac{{d^3 p}}{{dt^3 }} + \frac{{h^2 \chi }}{\tau }\left\{ { - \frac{{6\dot \rho ^3 }}{{\rho ^2 }}
  + \frac{{6\dot \rho \ddot \rho }}{\rho } - \frac{d^3 \rho }{dt^3 }}
  \right\}.
\end{array}
\end{equation}
Solutions (\ref{wave_sol}) satisfy the following dynamical system
\begin{equation}\label{dyns}
\begin{array}{l}
 \displaystyle U\frac{{dU}}{{d\omega }} = UW, \quad
 U\frac{{dZ}}{{d\omega }} = \gamma U + \xi Z + W(Z - U^2 ), \quad  U\frac{{dW}}{{d\omega }} = UR, \\
\displaystyle U\frac{{dR}}{{d\omega }} = \left( {bU^3 \left( {\chi  - U^2 } \right)} \right)^{ - 1} \{  - \kappa U^2  + \eta \xi \sigma W - 2\xi \sigma U^2 W + \chi \tau U^2W  - h\xi U^4 W +  \\
  + b\xi ^2 U^4 W - \tau U^4 W + \eta \sigma W^2  + \left( {\chi h - \sigma } \right)U^2 W^2  - hU^4 W^2  + b\xi U^4 W^2  - b\chi U^2 W^3  +  \\
  + bU^4 W^3  + \gamma \left( {2\xi \sigma U + h\xi U^3  - b\xi ^2 U^3  + \tau U^3  + 2\sigma UW} \right) + U^2 Z + h\xi ^2 U^2 Z -  \\
  - b\xi ^3 U^2 Z + \xi \tau U^2 Z + \left( { - \eta \sigma U + U^3 \left\{ {\sigma  + \chi h - 4b\chi W - hU^2  + b\xi U^2  + 4bWU^2 } \right\}}
  \right)R\},
 \end{array}
\end{equation}
where $b=h^2/\tau$, and quadrature
\[
U\frac{{dS}}{{d\omega }} =  - \left( {W + \xi } \right).
\]
The fixed point of this system has the coordinates
\begin{equation}\label{ospoint}
U_0  =  - D,Z_0  = \frac{{\kappa D^2 }}{{D^2  - 2\sigma \xi ^2
}},W_0  = 0,R_0  = 0.
\end{equation}

The conditions at which the linearized matrix $\hat M$
\begin{equation}\label{matrline}
 \hat J = \left(
{\begin{array}{*{20}c}
   0 & 0 & {a_1 } & 0  \\
   {a_2 } & {a_3 } & {a_4 } & 0  \\
   0 & 0 & 0 & {a_5 }  \\
   {a_6 } & {a_7 } & {a_8 } & {a_9 }  \\
\end{array}} \right) = \left( {\begin{array}{*{20}c}
   0 & 0 & { - D} & 0  \\
   \gamma  & \xi  & {Z_0  - D^2 } & 0  \\
   0 & 0 & 0 & { - D}  \\
   {a_6 } & {a_7 } & {a_8 } & {a_9 }  \\
\end{array}} \right),
\end{equation}

  $ \displaystyle a_6  = \frac{{\kappa \xi \left( { - 2\xi \sigma
+ D^2 \left( {h\xi  - b\xi ^2  + \tau } \right)} \right)}}{{\Delta
D\left( {2\xi ^2 \sigma  - D^2 } \right)}}$, $\displaystyle a_7  =
- \frac{{1 + h\xi ^2  - b\xi ^3  + \xi \tau }}{\Delta }$,

$\displaystyle a_8  =  - \frac{{\xi \sigma \left( {\eta  - 2Z_0 }
\right) - D^4 \left( {h\xi  - b\xi ^2  + \tau } \right) + D^2
\left( {\chi \tau - 2\xi \sigma } \right)}}{{D^2 \Delta }}$,

$\displaystyle a_9  = \frac{{\chi D^2 h - D^4 h + bD^4 \xi  + D^2
\sigma  - \eta \sigma }}{{D\Delta }}$, $\displaystyle \Delta  =
bD\left( {\chi  - D^2 } \right)$ admits the spectrum $(\pm
\Omega^2i;-\alpha_1;-\alpha_2)$ have the form
\begin{equation}\label{cond}
B_2  = \frac{{B_1 }}{{B_3 }} + B_0 \frac{{B_3 }}{{B_1
}},\;\;\;\;B_3 ^2  - 4B_0 \frac{{B_3 }}{{B_1 }} \ge 0,
\end{equation}
where $ B_3=- a_3 - a_9 $, $B_2=a_3 a_9 - a_5 a_8 $, $B_1 =
a_5\left( {a_3 a_8  - a_1 a_6 - a_4 a_7 } \right)$, $B_0=a_1 a_5
\left( a_3 a_6 - a_2 a_7 \right) $ are the coefficients of
characteristic polynomial for the matrix $\hat M$.

If we fix the parameters $\chi = \eta = 30$, $\xi = - 1.9$, $h =
1$, $\tau = 1$, $b=1$, $\sigma = 2.7$, then in the plane
$(D^2,\kappa)$  Eq.~(\ref{cond}) defines the NSC. Crossing this
curve in the point $A (2.2852;3.7)$, one can observe the
appearance of the limit cycle at $D^2 \geq 2.2852$.

\begin{figure}[t]
\includegraphics[width=7.5 cm, height=6 cm ]{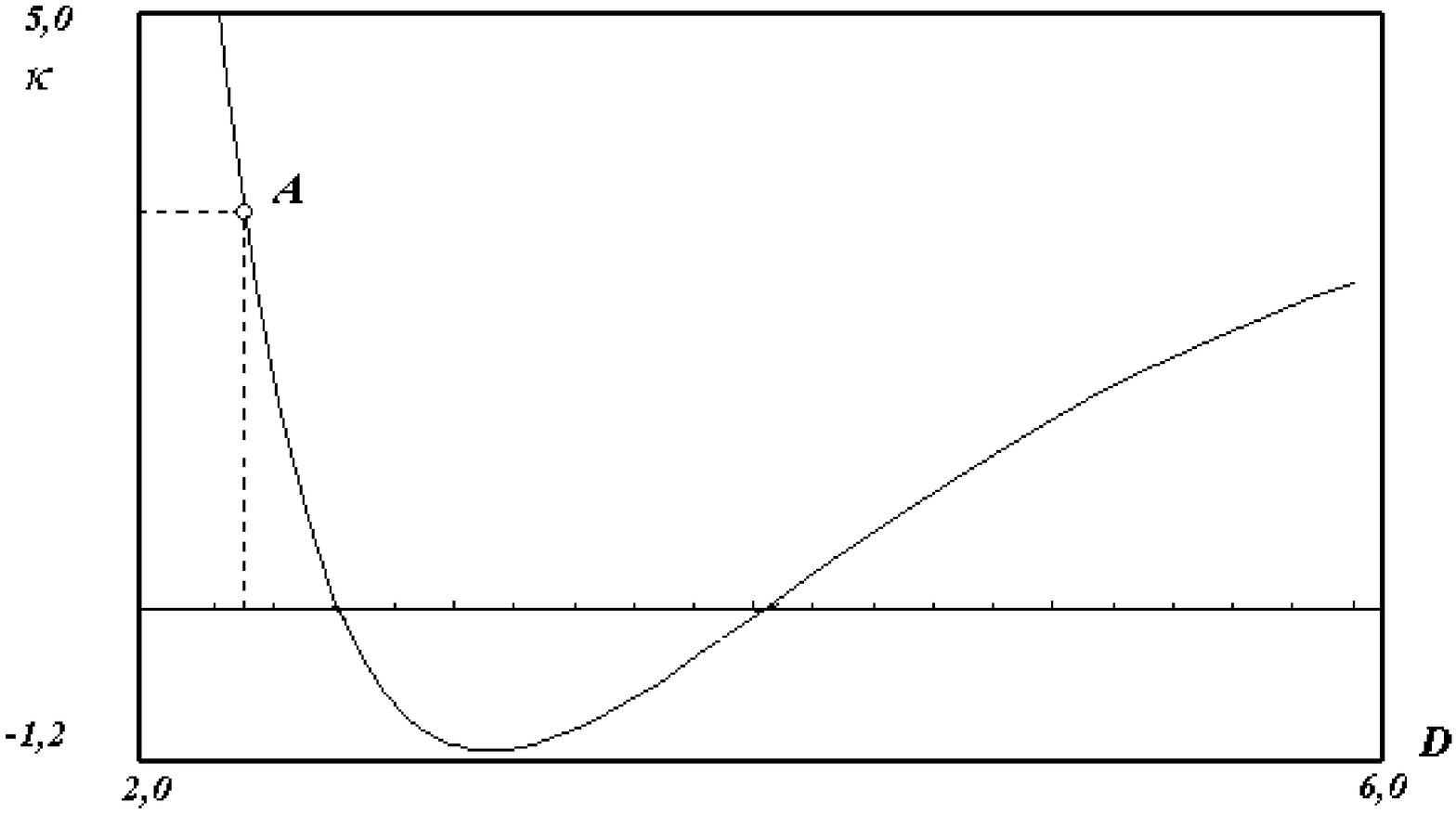}
\hfill
\includegraphics[width=7.5 cm, height=6 cm ]{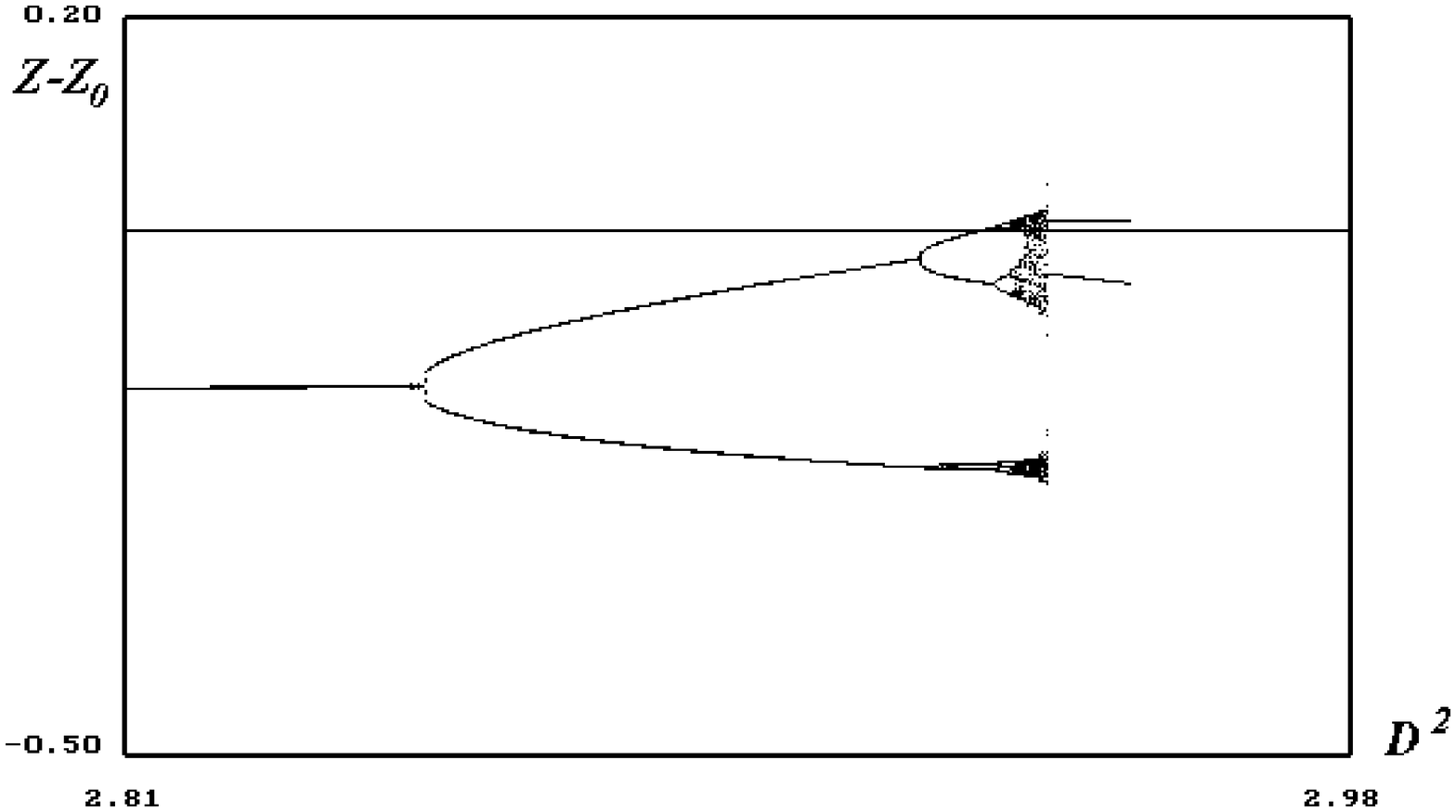}
 \centerline{a) \hspace{8cm} b)}
\caption{a) Neutral stability curve in the plane  $(D^2;\kappa)$.
b) The bifurcation Poincar\'{e} diagram at increasing
$D^2$}\label{deriv_3}
\end{figure}

\begin{figure}[th]
\includegraphics[width=7.5 cm, height=6 cm ]{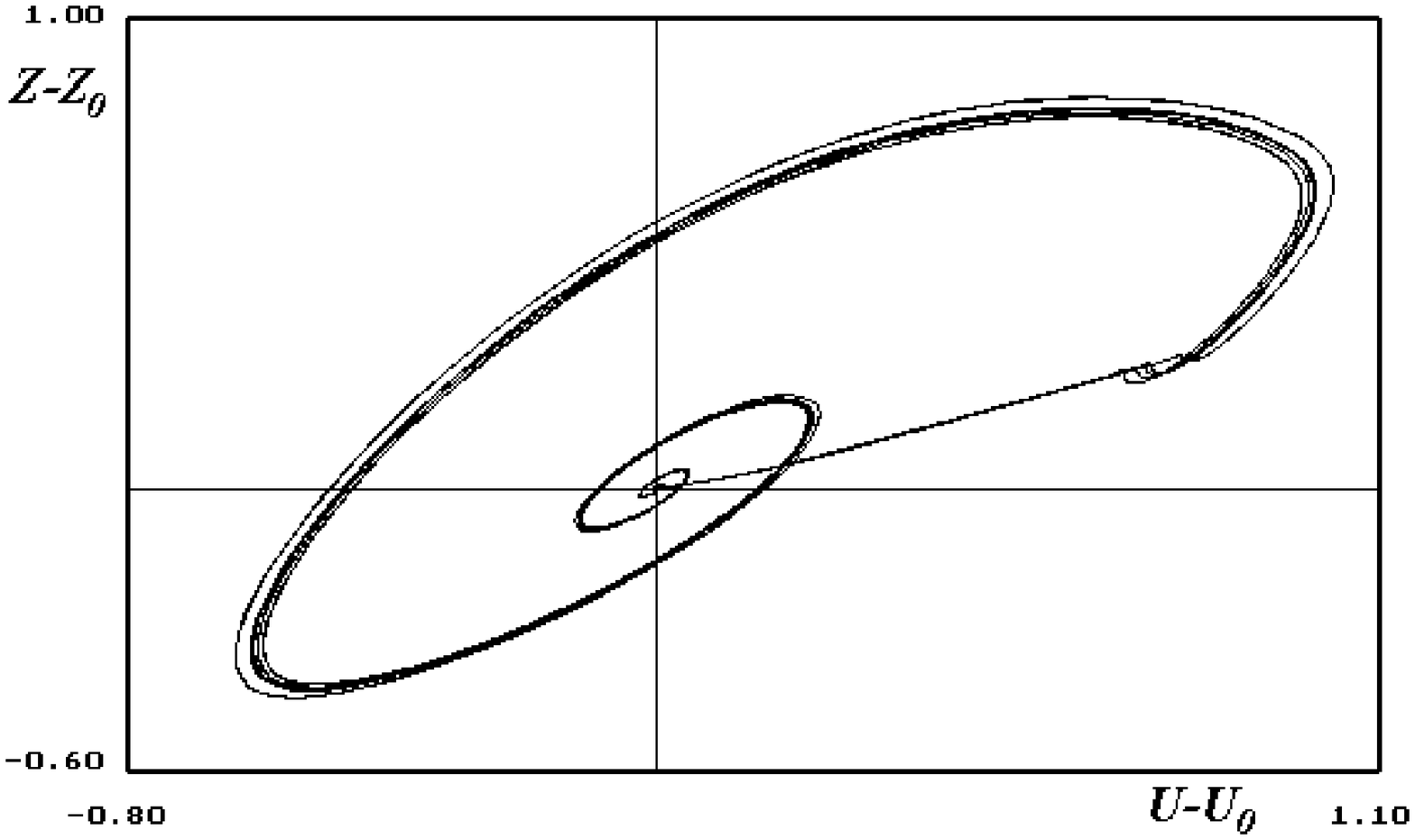}
\hfill
\includegraphics[width=7.5 cm, height=6 cm ]{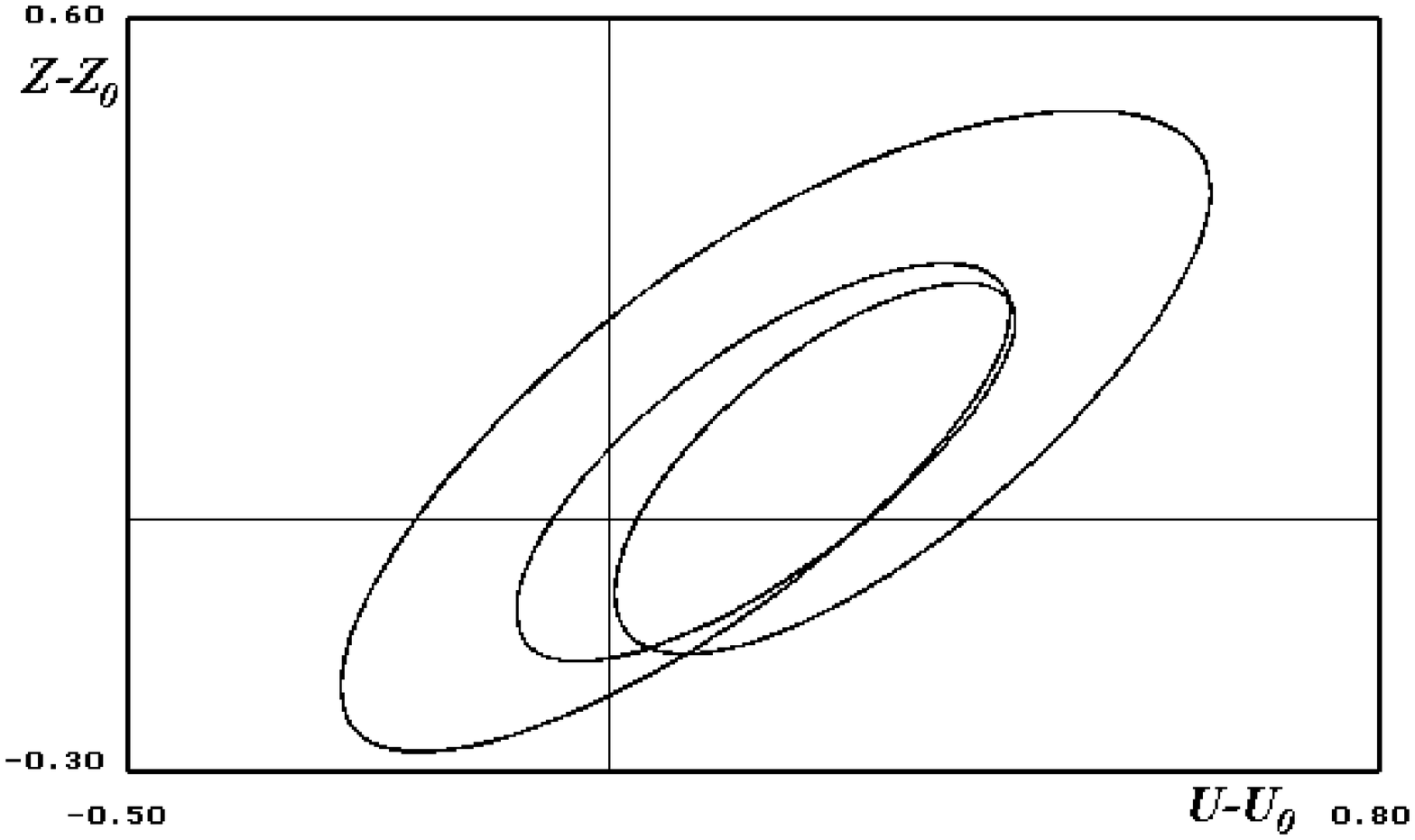}
 \centerline{a) \hspace{8cm} b)}
\caption{Phase portraits of separated trajectories derived at
 $D^2=2.722,\,\kappa=3.7,\,b=1$ and different initial conditions.}\label{deriv3_phase}
\end{figure}

In the Poincar\'{e} diagram depicted at increasing $D^2$
(Fig.\ref{deriv_3}) we can identify the moments of several period
doubling bifurcations leading to the chaotic attractor creation.
But the chaotic attractor existing at a short interval of
parameter $D^2$ is destroyed. Instead of it  in the phase space of
system (\ref{dyns}) the complicated periodic trajectory resembling
to a loop (Fig.~\ref{deriv3_phase}a) appears.

\begin{figure}[th]
\includegraphics[width=7.5 cm, height=6 cm ]{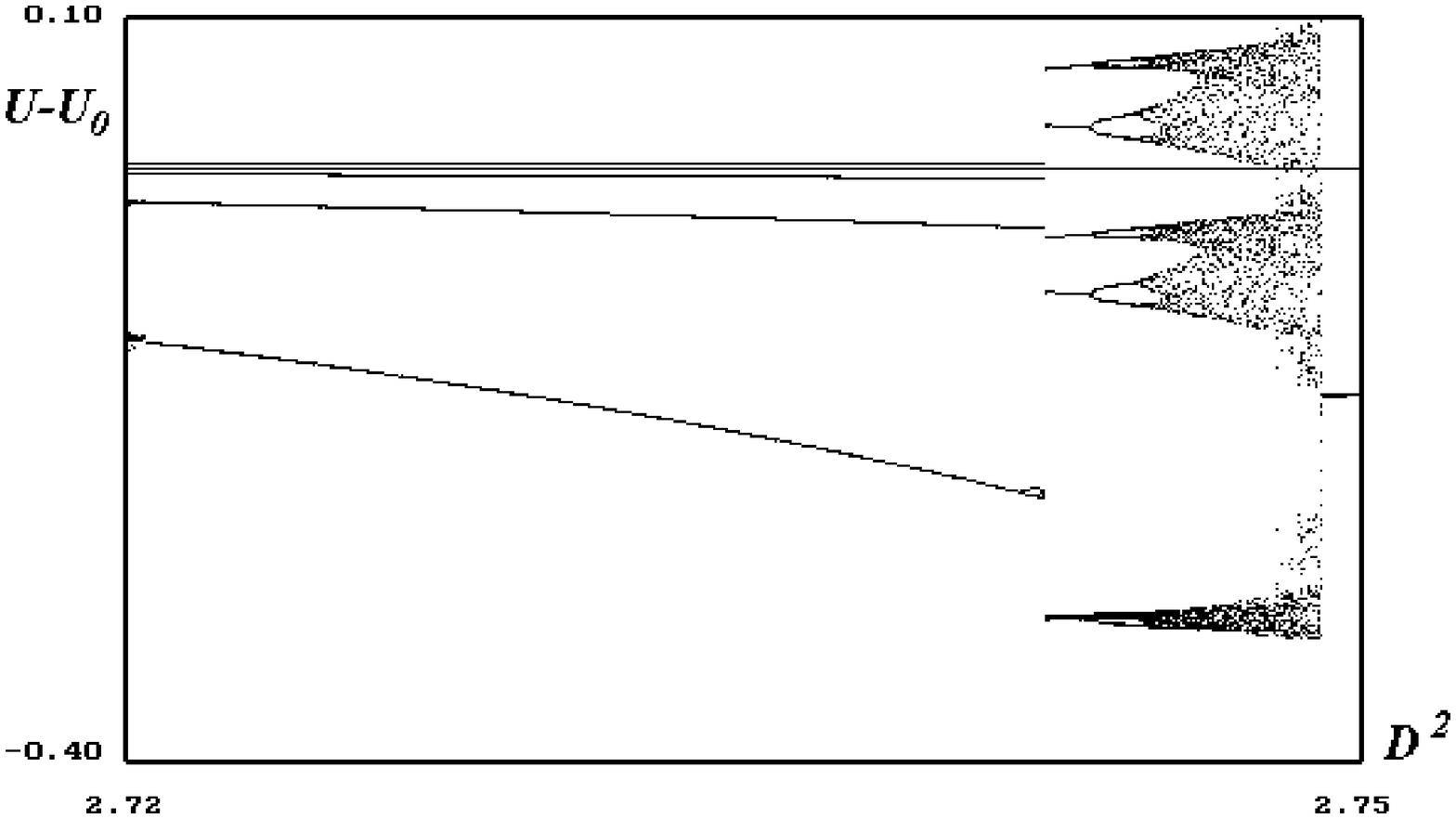}
\hfill
\includegraphics[width=7.5 cm, height=6 cm ]{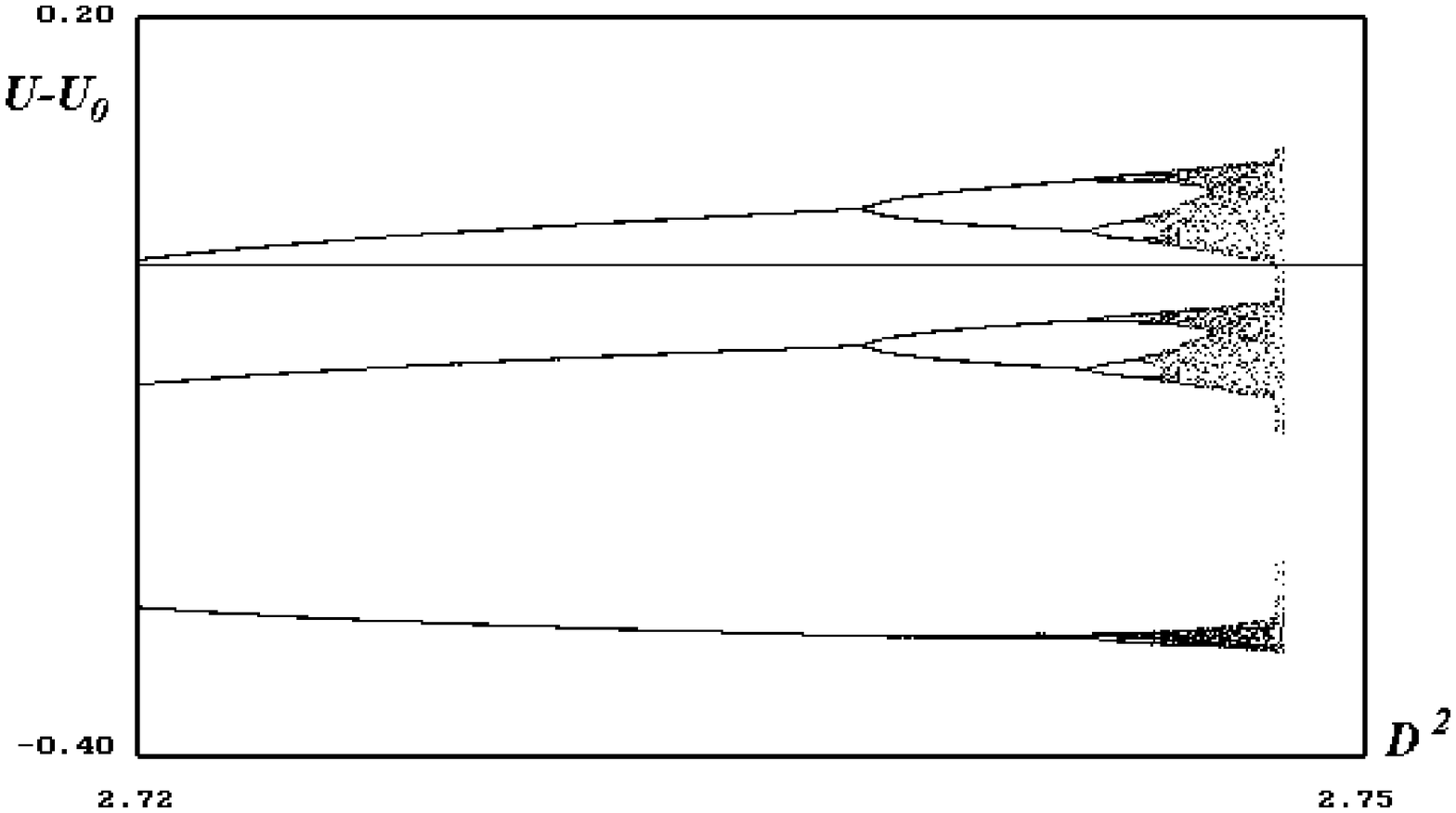}
 \centerline{a) \hspace{8cm} b)}
\caption{ The bifurcation Poincar\'{e} diagram of development of
separated regime at   increasing $D^2$ (a) and decreasing $D^2$.
Here $b=1$.}\label{deriv3_diag}
\end{figure}

Consider also the development of oscillating regimes whose basins
of attraction are separated from the basin of attraction of the
main limit cycle. Integrating dynamical system (\ref{dyns}) from
initial conditions $\left( {0;0;0;0.01} \right)$ at $D^2=2.722$,
we see that the phase space of the system, in addition to the main
limit cycle, contains the complicated trajectory
(Fig.~\ref{deriv3_phase},a) which can be regarded as a hidden
attractor. From the analysis of  Poincar\'{e} diagram
(Fig.~\ref{deriv3_diag}a) it follows that the system weakly
responds to  the growing of the parameter $D^2$ until
$D^2=2.7445$. When $D^2>2.7445$, the system jumps  to another type
of oscillations followed by  chaotic regime creation.

If we plot the Poincar\'{e} diagram  at decreasing $D^2$
(Fig.~\ref{deriv3_diag}b) starting from the chaotic attractor,
then we observe the periodic trajectory (Fig.\ref{deriv3_phase}b)
that differs from the initial regime (Fig.\ref{deriv3_phase}a).
Note that the periodic trajectory from Fig.~\ref{deriv3_phase}b
can be revealed directly by the integration from the initial
conditions $\left( 0;0;0;0.1 \right)$.

\section{DES with physical nonlinearity and second derivatives}\label{sec6}

Till now we dealt with the models without physical nonlinearity.
Generalizing the previous models in this direction, we obtain the
following model \cite{ds_rep_07}
\begin{eqnarray}
  \sigma \chi
\rho ^{ - 2} \left[ { - \rho _{xx} \left( {1 + a} \right) + \rho
_x^2 \rho ^{ - 1} \left( {1 - na} \right)} \right]
 + \nonumber \\ +h\chi \rho ^{ - 2}  [  - \ddot \rho \left( {1 + a} \right) + \vspace{0.25 cm}
 2\dot \rho ^2 \rho ^{ - 1} \left( {1 - 0.5a(n - 1)} \right)
  +\tau h^{ - 1} \dot \rho \left( {1 + a} \right) ]
 + \kappa \rho ^n  = p + \vspace{0.25 cm} \label{vizn11}\\
 +\tau \dot p - h\ddot p - \sigma \left( {p_{xx}  + \rho _x p_x \rho ^{ - 1} }
 \right),\quad a = \delta n\rho ^{n + 1}.\nonumber
\end{eqnarray}
 Properties of solutions to  system (\ref{vizn11}) can be found out using the symmetry of the
system with respect to the Galilei group \cite{symm}. One can be
persuaded by direct verification that system (\ref{vizn11}) allows
operator
\[
\hat X = \frac{1}{{2\xi }}\frac{\partial }{{\partial t}} +
t\frac{\partial }{{\partial x}} + \frac{\partial }{{\partial u}}.
\]
Let us construct an anzatz with its invariants
\begin{equation}\label{sol}
\begin{array}{c}
\rho  = R(\omega ), \quad  p = P(\omega ), \quad u = 2\xi t +
U(\omega ), \quad \omega  = x - \xi t^2,
\end{array}
\end{equation}
where parameter $\xi$ is proportional to acceleration of the wave
front. Substitution (\ref{sol}) into the system yields the
following  quadrature
\[ UR = C =\mbox{const}
\] and the dynamical system
 \begin{equation}\label{dynsys}
 \begin{array}{c}
  \displaystyle  R^\prime = W, \quad P^\prime  = \gamma R^m  - 2\xi R + {{C^2 } \over {R^2
  }}W,\vspace{0.1 cm}\\
\displaystyle  W^\prime  =  - (\kappa R^{n + 3}  - PR^3
  - P'R^2 C \tau - hP'C^2 W + P'R^2 \sigma W + \vspace{0.1 cm}\\+ \gamma m R^{2 + m} \sigma W
     + \chi L \tau C W + \gamma hmR^{m} C^2 W + h\chi L (CWR^{-1})^2 - \vspace{0.1 cm}\\
     - 2C^2 \sigma W^2  + \chi M\sigma W^2  - 2C^4 h R^{-2} W^2  +\vspace{0.1 cm}\\
    + 2h\chi NC^2R^{-2} W^2  - 2R^3 \sigma W\xi  - 2hR C^2 W\xi
    )\times\vspace{0.1 cm}\\
    ((C^2  - \chi L)R(\sigma  + hC^2R^{-2} ))^{ - 1},
    \end{array}
\end{equation}
 where $\displaystyle  ( \cdot )' = {d \over {d\omega }}\left(
\cdot  \right),L = 1 + a,M = 1 - an,N = 1 - 0.5a(n - 1),a = \delta
nR^{n + 1} .$

The single isolated equilibrium (neglecting the trivial) point has
the following coordinates
\begin{equation}\label{crp1}
R_0  = \left( {{{2\xi } \over \gamma }}
\right)^{{\raise0.7ex\hbox{$1$} \!\mathord{\left/
 {\vphantom {1 {n - 1}}}\right.\kern-\nulldelimiterspace}
\!\lower0.7ex\hbox{${m - 1}$}}} ,\,P_0  = \kappa R_0^n ,\,W_0  =
0.
\end{equation}
 In this point the linearized matrix $\hat M$ has the form
\begin{equation}\label{lins}
\hat M=\left( {\begin{array}{*{20}c}
   0 & 0 & 1  \\
   {a_1 } & 0 & {a_2 }  \\
   {a_3 } & {a_4 } & {a_5 }  \\
\end{array}} \right),
\end{equation}
where
\begin{eqnarray*}
a_1  = 2\xi (n - 1), \quad a_2  = C^2 R_0^{-2},\\
 a_3  = (2C^3 h\left[ {C^2  - \chi L} \right]\tau
 \left[ {\gamma R_0^m  - 2\xi R_0 } \right]R_0^{ - 2} +  \\ + C\chi (n + 1)(L - 1)\tau \Delta  -
  C\left[ {C^2  - \chi L} \right]\tau \Delta  -  \\
  - \left[ {C^2  - \chi L} \right]
 \left( {C^2 hR_0^{ - 2}  + \sigma } \right)
 \left( {\kappa nR_0^{1 + n}  - C\tau \left( {\gamma (2 + m)R_0^m  - 6\xi R_0 } \right)} \right))/\Delta ^2,
 \\
\displaystyle a_4  = R_0^2 \Delta ^{ - 1} , \\ a_5  = \frac{{C^2
\gamma h\left( {nR_0^n  - R_0^m } \right) - C^3 \tau  + C\chi
L\tau  + R_0^2 \sigma \left( {\gamma \left[ {R_0^m  + nR_0^n }
\right] - 4R_0 \xi } \right)}}{{R_0 \Delta }}, \\
\Delta  = \left( {C^2 - \chi L} \right)\left( {C^2 hR_0^{ - 2}  +
\sigma } \right).
\end{eqnarray*}

The NSC for system (\ref{dynsys}) has the following form
\begin{equation}\label{nsc} G\left( {\xi ,\sigma ,n,h,\tau ,\kappa ,\chi } \right)
\equiv a_5 \left( {a_3  + a_2 a_4 } \right) + a_1 a_4  = 0.
\end{equation}

Let us make the values of parameters fixed as follows:
\begin{eqnarray*}
\gamma=1,\quad \chi=10,\quad C=-2.8, \quad \sigma = 0.2,\,
\tau=1.1,\, h=3.2,\, \delta=1.4, \, n=m=3.2. \end{eqnarray*}

Condition (\ref{nsc}) allows us to find numerically the value of
$\xi_0=0.157$ corresponding to birth of the limit cycle.

Let us consider in more detail the influence on the revealed
regimes of parameters $n$ and $\delta$ changes, which determine
nonlinearity of the medium in the dynamic equation of state. Let
us make the value of parameter $\xi=0.35$ fixed, in case of which
there is a limit cycle with period $2T$ in the space of the
system.

\begin{figure}[h]
\includegraphics[width=7.5 cm, height=6 cm ]{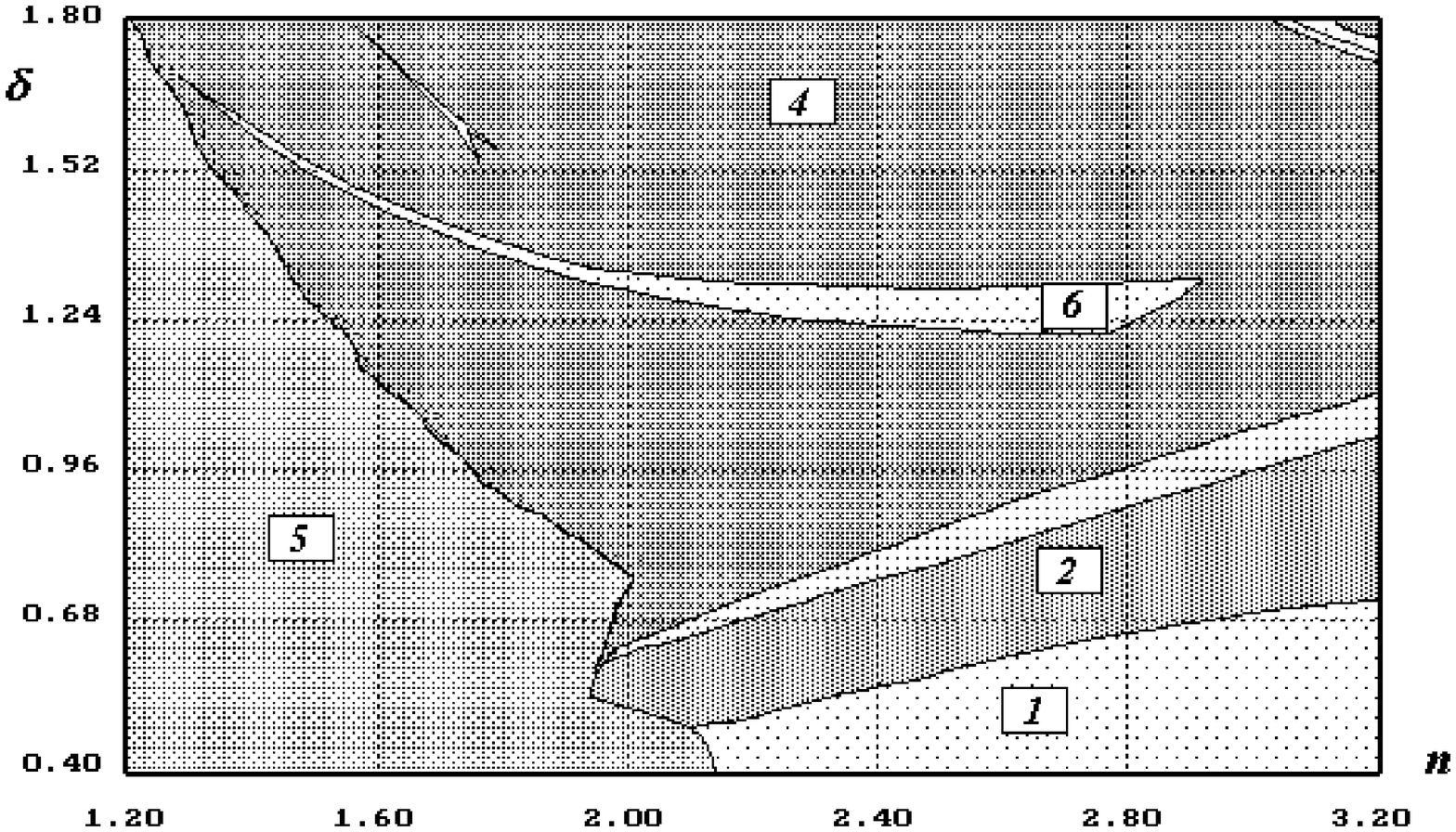}
\hfill
\includegraphics[width=7.5 cm, height=6 cm ]{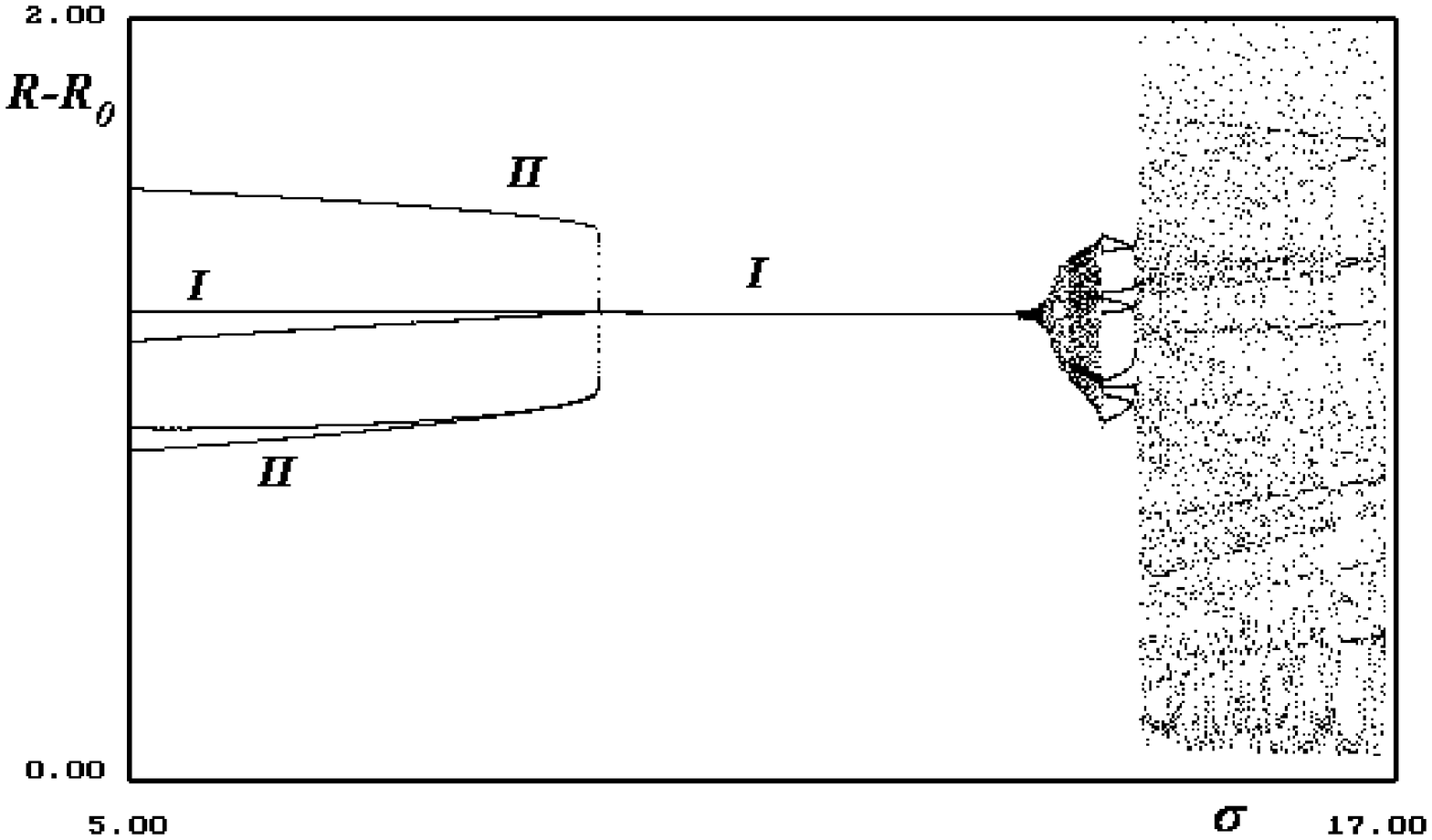}
 \centerline{a \hspace{8cm} b}
\caption{ a) Two-parametric bifurcation diagram in case of
$\kappa=2$ (for other values of parameters and conventional
symbols see Fig.\ref{phas_soliton}; b) Poincar\'{e} bifurcation
diagram for development of the torus in case of $\gamma=1$,
$\chi=10$, $C=-2.8$, $\tau=1.1$, $\kappa=0.9$, $h=3.2$,
$\delta=0.4$, $\xi=0.35$, $n=m=3.2$ and increasing $\sigma$, where
graph I is the basic limit cycle, graph II -- complicated periodic
trajectory with separated region of attraction. }\label{figds3}
\end{figure}

 The diagram
reveals some peculiarities of system's (\ref{dynsys}) behavior. In
particular, we would like to pay attention to the  presence of a
''special'' $\,$ point in the parameter plane, surrounded by four
different types of solutions. One can also see the ''windows'' of
periodicity (area 6) among the chaotic area. To find out the
structure of phase space in more detail near  area 6 of
Fig.\ref{figds3}a, let us plot a one-parametric Poincar\'{e}
diagram (Fig.\ref{figds3}b) for $\delta=1.4$ and a decrease of
parameter $n$.

In case of $n$ close to 2, abrupt reconstruction of the chaotic
attractor structure can be  observed, which is probably caused by
the  interaction of two (or more) co-existing attractors of a
dynamic system. In case of $n\approx 1.4$ the chaotic trajectory
is localized in a more narrow area of  phase space of  system
(\ref{dynsys}), stipulating the appearance of a specific window of
periodicity with a decrease of $n$. Analysis of a two-parametric
bifurcation diagram for the value of parameter $\kappa=2$
(Fig.\ref{figds3}a) shows that the area of existence of the
chaotic attractor increases and the windows of regular intervals
in case of the increasing $\kappa$ shift towards higher values of
the nonlinearity parameter $n$.

A crucially different set of bifurcations is observed in  case of
a change of  parameter $\sigma$.

Let us fix the values of parameters $\gamma=1$, $\chi=10$,
$C=-2.8$, $\tau=1.1$, $\kappa=0.9$, $h=3.2$, $\xi=0.35$, $n=m=3.2$
and $\delta=0.4$. Integrating  system (\ref{dynsys}) with initial
data $(0,\, 0,\, 0.01)$  and  $\sigma =5$ within phase space near
the equilibrium point, in addition to the limit cycle,  other
periodic trajectory has been found with a separated pool of
attraction (development of this regime with increasing of $\sigma
$ is presented in Fig.\ref{figds3}b graph II).

The presence of such a regime leads to the thought of the
existence of quasi-periodic regimes. To look for such a regime let
us plot a bifurcation diagram of Poincar\'{e} for
   development of basic limit cycle in case of increasing
 parameter $\sigma $ (Fig.\ref{figds3}b graph I).

\begin{figure}[h]
\includegraphics[width=7.5 cm, height=6 cm ]{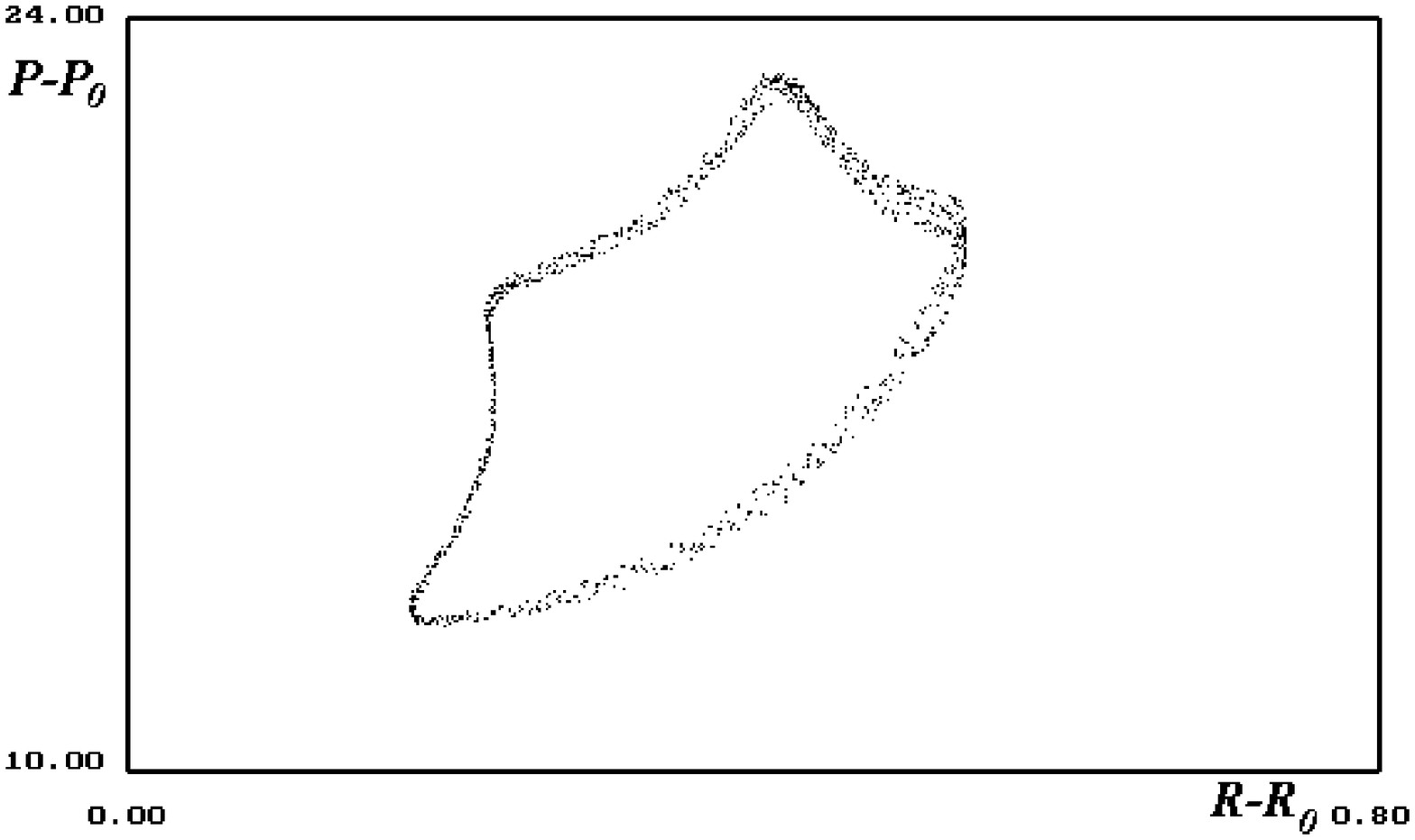}
\hfill
\includegraphics[width=7.5 cm, height=6 cm ]{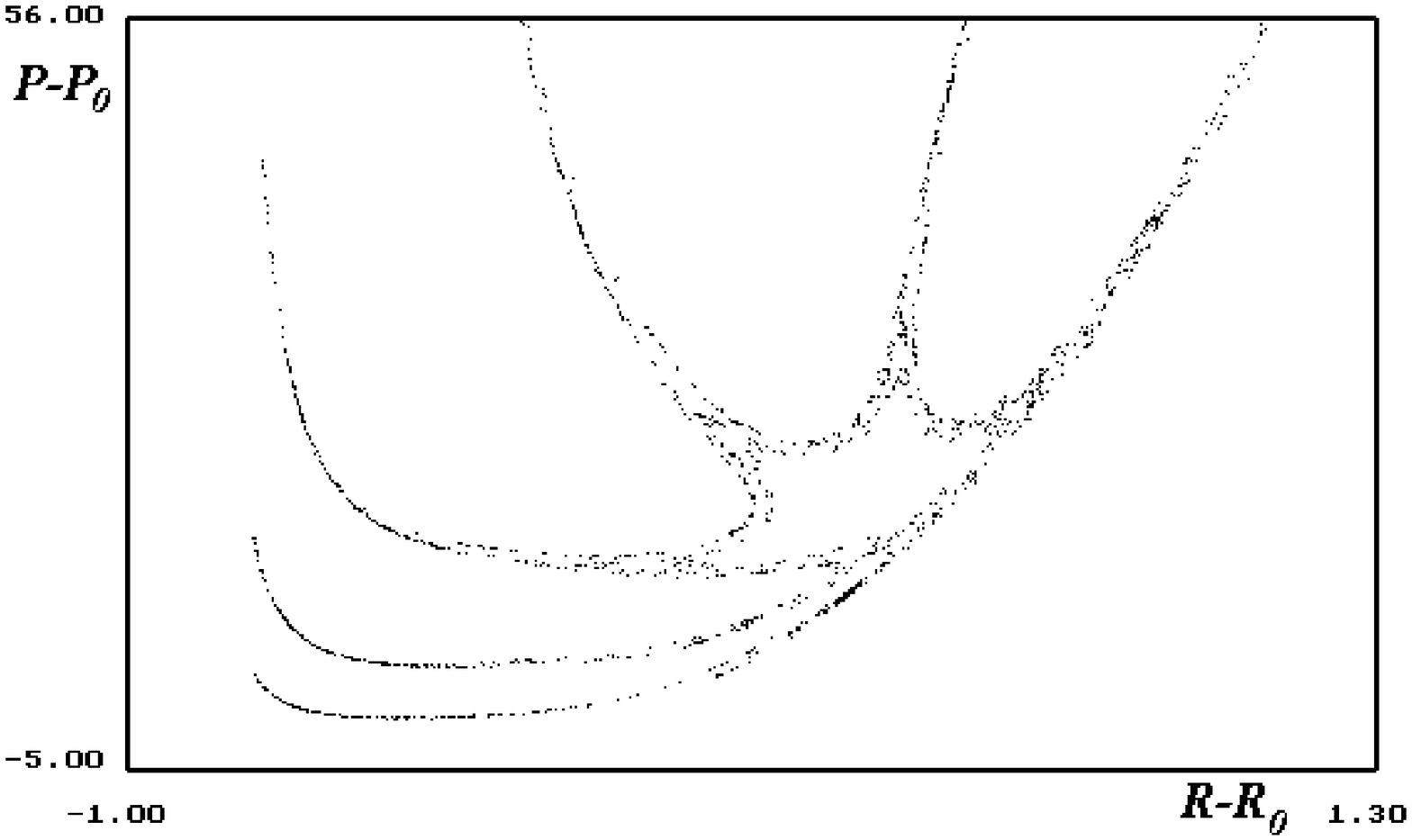}
 \centerline{a \hspace{8cm} b}
\caption{a) The Poincar\'{e} cross-section of the torus surface in
case of $\sigma=14$ b) The Poincar\'{e} cross-section of a chaotic
attractor in case of $\sigma=14.6$.  Fixed parameters $\gamma=1$,
$\chi=10$, $C=-2.8$, $\tau=1.1$, $\kappa=0.9$, $h=3.2$,
$\delta=0.4$, $\xi=0.35$, $n=m=3.2$.  }\label{figds4}
\end{figure}

 Another bifurcation, leading to the appearance of the toroidal
surface, has been discovered in this system. An intersection of
the toroidal attractor with the plane $y_3=0$ forms a closed
curve, shown in Fig.\ref{figds4}a. A further increase of parameter
$\sigma $ causes the synchronization of tore frequencies, and
finally an abrupt increase of vibrations amplitude, which shows
the creation of a crucial new dynamical behavior. To clarify the
character of the produced regime, let us analyze the Poincar\'{e}
section for the case of $\sigma=14.6$ (Fig.\ref{figds4}b). The
plotted
       cross-section is specific for chaotic attractor, which provides reasons for statements
          on the existence of bifurcation of a quasi-periodic regime with a producing chaotic attractor.

It turned out that system (\ref{dynsys}) provides another type of
chaotic attractor creation, namely, intermittency. Let us fix
$\gamma=1$, $\chi=50$, $C=-1.5$, $\tau=0.1$, $\kappa=1.9$,
$\sigma=0.2$, $h=0.9$, $\xi=0.18$.

\begin{figure}[th]
\includegraphics[width=7.5 cm, height=6 cm ]{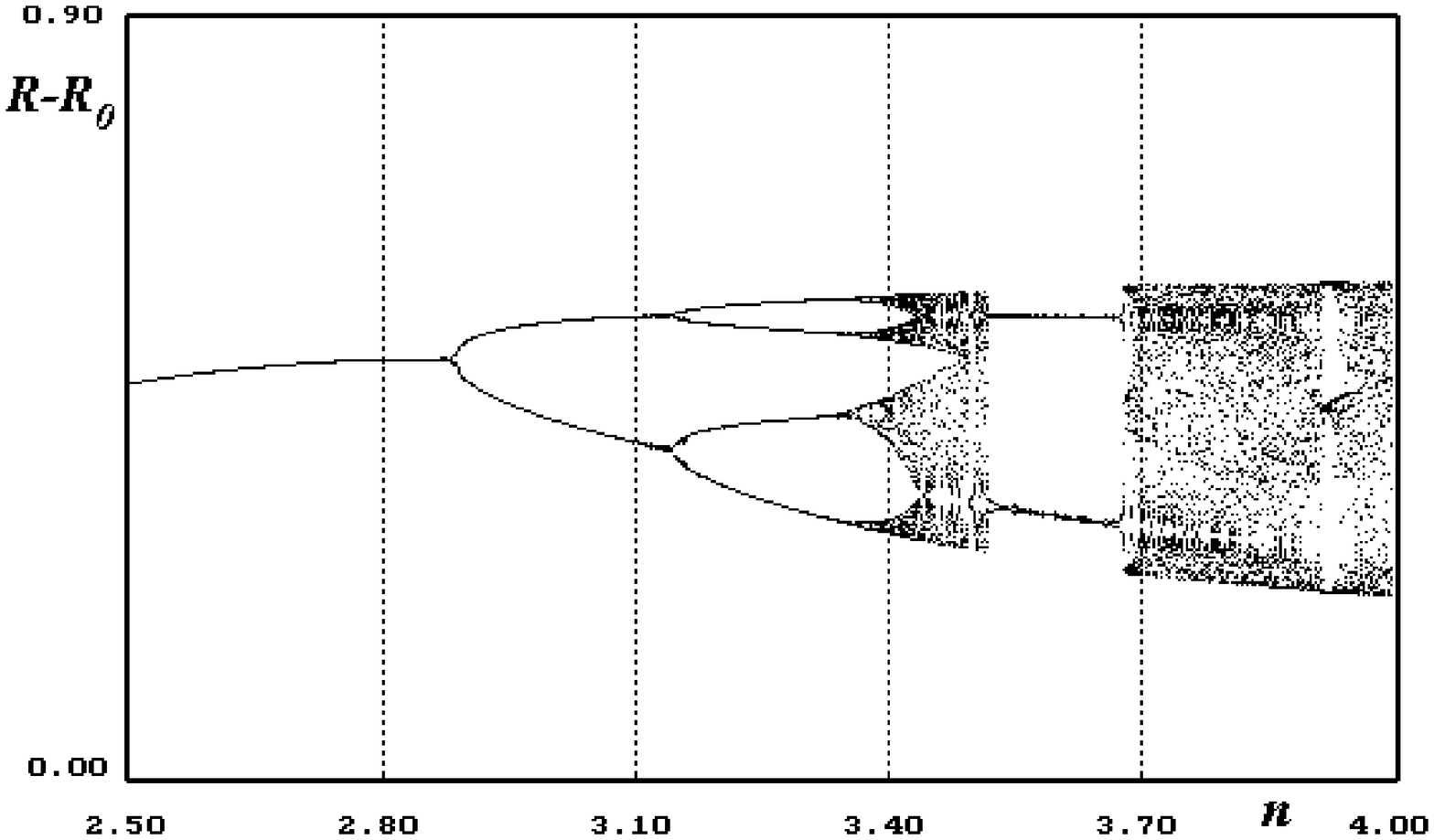}
\hfill
\includegraphics[width=7.5 cm, height=6 cm ]{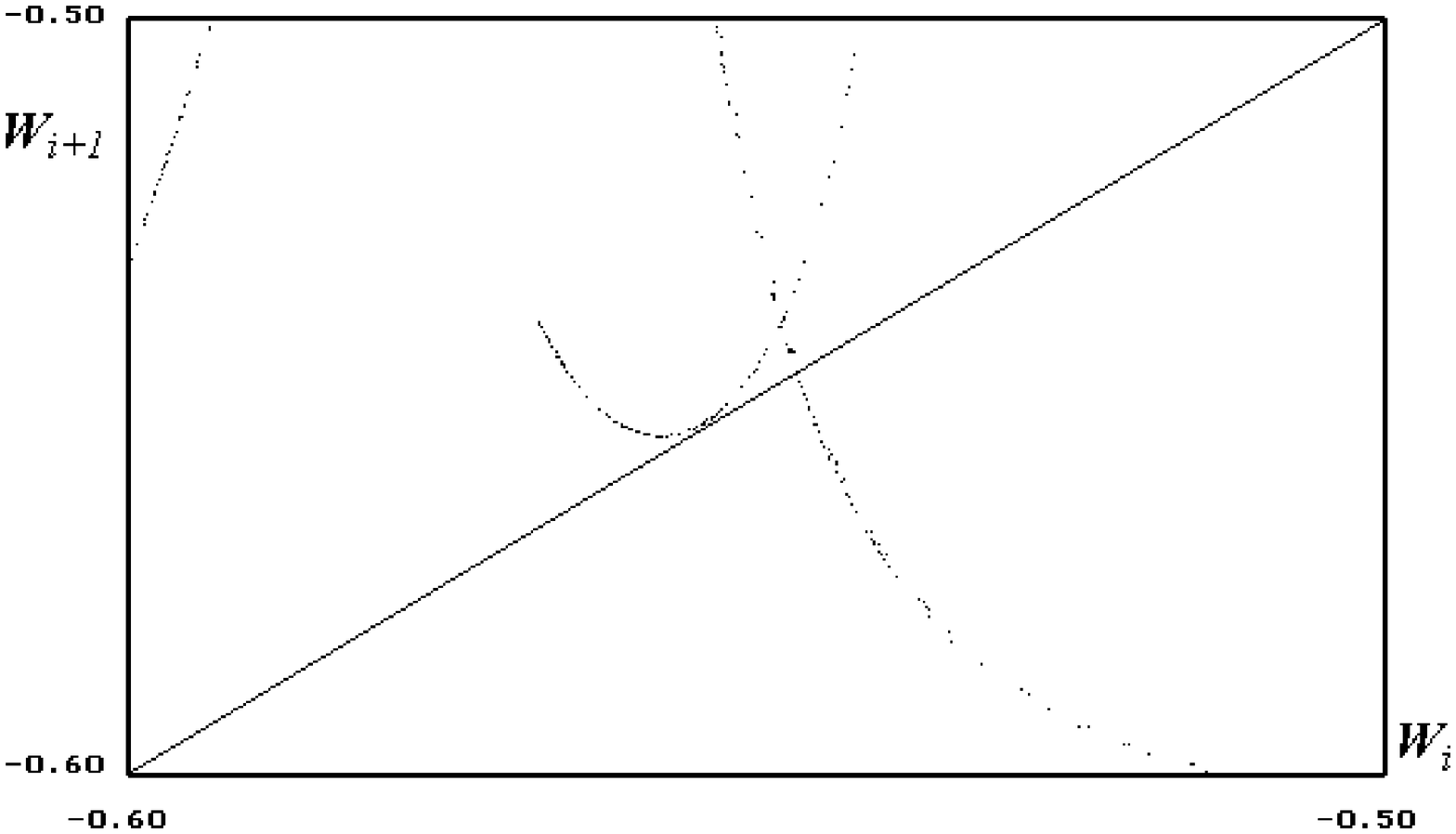}
 \centerline{a \hspace{8cm} b}
\caption{ a) The bifurcation diagram at increasing $n$. b) The
graph of dependence $ W_{i+1}$ vs $ W_{i}$ at $n=4.25$. The fixed
values of parameters $\gamma=1.49,
\,\chi=50,\,C=-1.5,\,\tau=0.1,\,\kappa=1.9,\,\sigma=0.2,\,h=0.9,\,\xi=0.18$,
$\delta=0,8$.}\label{nonlin}
\end{figure}

Plotting the Poincar\'{e} bifurcation diagram
(Fig.~\ref{nonlin}a), we see that a limit cycle undergoes several
period doubling bifurcations resulting in the chaotic attractor
creation. But the development of chaotic attractor is  interrupted
suddenly and new complicated periodic trajectory appears which
bifurcates in chaotic attractor as well at increasing $n$.
Considering the hereditary sequences (Fig.\ref{nonlin}b) for
chaotic trajectories, we found that the graph of the map
$W_{i+1}=f(W_i)$ is close to the bissectrice at $n=4.25$. As in
the case  with the Lorentz system, existence  of narrow passage
leads to the alternation of the chaotic and regular behavior of
the system trajectories.

\section{Conclusions}\label{sec_concl}

Finally, we have studied the hierarchical sequences of the
mathematical models for non-equilibrium media. Analyzing the wave
fields in such  media we have shown that derived models possess
the wide set of localized wave regimes. In particular, the models
with relaxation admit the periodic, multiperiodic, chaotic
solutions. Spatially nonlocal models have in addition
quasiperiodic and solitary wave solutions. All the models
demonstrate the most bifurcations and scenarios of chaotic regimes
creation.

From the other hand, identifying internal variables with
parameters undergoing fluctuations, one can consider these
investigations as the problem on the dissipative structures
creation under the influence of noise.



%
%

%


\bibliography{danylenko}

\end{document}